%% file: main.tex
\documentclass[manuscript]{acmart}

\AtBeginDocument{%
  \providecommand\BibTeX{{%
    \normalfont B\kern-0.5em{\scshape i\kern-0.25em b}\kern-0.8em\TeX}}}

\setcopyright{rightsretained}

\copyrightyear{2026} 
\acmYear{2026}
\setcopyright{rightsretained}
\acmDOI{10.xxxx/xxxxxxx.xxxxxxx} 
\acmISBN{978-1-4503-8655-5/21/10}

\usepackage{multirow}
\usepackage{subcaption}
\usepackage{caption}
\usepackage{threeparttable}
\usepackage{tabularx}
\usepackage{pifont}
\usepackage{xcolor}
\usepackage{makecell}
\usepackage{colortbl}
\newcommand{\cmark}{{\color{green!70!black}\ding{51}}}

\usepackage[utf8]{inputenc}
\usepackage{xcolor}
\usepackage[most]{tcolorbox}

\tcbuselibrary{breakable}

\begin{document}

\title[CareerPooler: AI-Powered Metaphorical Pool Simulation]{CareerPooler: AI-Powered Metaphorical Pool Simulation Improves Experience and Outcomes in Career Exploration}

\author{Ziyi Wang}
\affiliation{%
  \institution{University of Maryland, College Park}
  \country{USA}}

  \author{Ziwen Zeng}
\affiliation{%
  \institution{iHeartMedia}
  \country{USA}}

  \author{Yuan Li}
\affiliation{%
  \institution{University of Alabama}
  \country{USA}}

  \author{Zijian Ding}
\affiliation{%
  \institution{University of Maryland, College Park}
  \country{USA}}

\renewcommand{\shortauthors}{Wang, et al.}

\begin{abstract}

Career exploration is uncertain, requiring decisions with limited information and unpredictable outcomes. While generative AI offers new opportunities for career guidance, most systems rely on linear chat interfaces that produce overly comprehensive and idealized suggestions, overlooking the non-linear and effortful nature of real-world trajectories. We present CareerPooler, a generative AI–powered system that employs a pool-table metaphor to simulate career development as a spatial and narrative interaction. Users strike balls representing milestones, skills, and random events, where hints, collisions, and rebounds embody decision-making under uncertainty. In a within-subjects study with 24 participants, CareerPooler significantly improved engagement, satisfaction, and career clarity compared to a chatbot baseline. Qualitative findings indicate that spatial-narrative interaction promotes experience-based learning, resilience through setbacks, and reduced psychological burden. Our findings contribute to the design of AI-assisted career exploration systems and, more broadly, suggest that visually grounded metaphorical interactions can make generative systems more engaging and satisfying.
\end{abstract}

 



\begin{teaserfigure}
 \centering
 \includegraphics[width=1\textwidth]{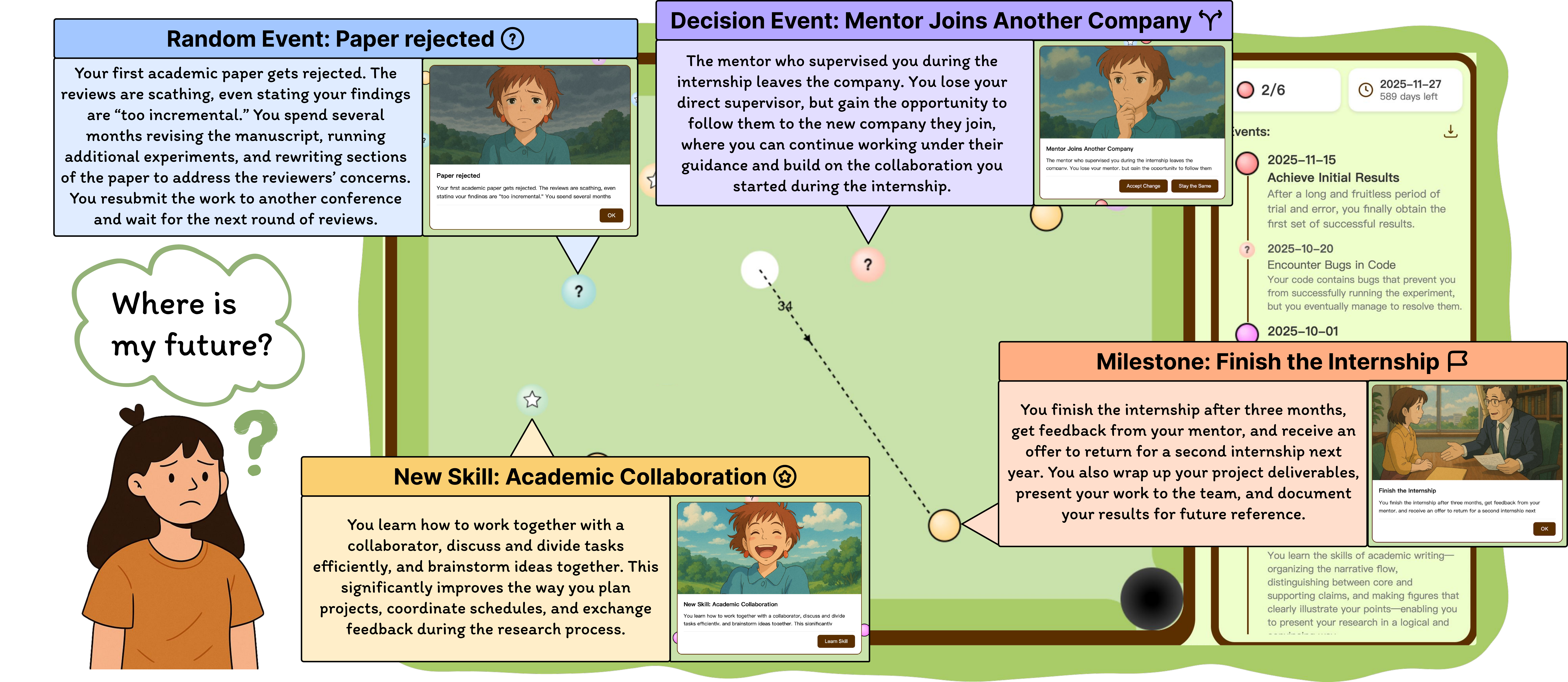}
   \caption{CareerPooler experience scenario: while facing uncertainty about future career, the user explores her career journey through CareerPooler. Each ball represents different types of events she might experience along her journey.}
   \Description{CareerPooler Experience Scenario}
   \label{fig:teaser}
\end{teaserfigure}

\maketitle

\def \RQA {\textbf{RQ1 - Experience}: How does a spatially gamified AI career simulation influence user experience compared to AI chatbots?}

\def \RQB {\textbf{RQ2 - Outcome}: How does such a system influence career exploration outcomes compared to AI chatbots?}

\def \RQC {\textbf{RQ3 - AI Awareness}: How do individuals perceive AI’s impact on careers, and how do they prepare for it?}

\input{sections/1-introduction}

\input{sections/2-relatedWork}

\input{sections/3-systemDesign}

\input{sections/4-studyDesign}

\input{sections/5-results}

\input{sections/6-discussion_conclusion}

\input{sections/7-appendix}

\end{document}

%% file: sections/1-introduction.tex
\section{Introduction}
\label{sec:intro}

Career exploration, defined as the deliberate gathering of information that helps individuals advance in their careers~\cite{zikic2006job, jiang2019career}, is critical for creating coherent career plans, finding meaningful work, and adapting to life transitions~\cite{hirschi2009career}. Yet exploring career options often feels like navigating a maze: individuals must make decisions under uncertainty, with limited knowledge and evolving constraints~\cite{jiang2019career}. These challenges are compounded by structural forces such as economic instability and technological disruption. Generative AI, for instance, is automating entry-level positions in software development and customer service~\cite{handa_which_2025, shaoFutureWorkAI2025}, while elevated unemployment among recent graduates intensifies the pressure on those beginning or reorienting their careers~\cite{martinEducatedUnemployed2025}.

A growing number of AI-based tools have been introduced to support career exploration, but most rely on linear chat interfaces that produce overly comprehensive and idealized suggestions. Rule-based chatbots, mentorship platforms, and online communities~\cite{umbelinoEmergentUnderstandingMentor2021, wolfbauerReboWorkReflecting2023, tomprouCareerMentoringOnline2019} provide helpful guidance but often require extensive user input and are more effective for individuals who already have a clear sense of direction. More recently, creative approaches such as ``letters to my future self''~\cite{jeonLettersFutureSelf2025} and time-travel metaphors~\cite{pataranutapornFutureYouConversation2024, du2024careersim} have been introduced to encourage reflection. However, these tend to generate static snapshots of idealized futures. Because real-world career development is inherently iterative and path-dependent, where earlier decisions constrain later options and setbacks reshape trajectories, such static projections fail to capture the dynamic, non-linear nature of the process. Support systems that are more flexible, enjoyable, and encourage people to try new possibilities are deeply needed.


In this work, we present \textit{CareerPooler}, an interactive system for career exploration that adopts the metaphor of a pool table as its core interaction paradigm. Research on uncertainty in games suggests that the struggle to master uncertainty is central to engagement~\cite{costikyan2013uncertainty}, while studies connecting uncertainty with curiosity theory demonstrate that different types of uncertainty can motivate exploration and accommodate varying curiosity levels~\cite{to2016integrating}. Drawing on these insights and on spatial reasoning, a fundamental human capacity for understanding and anticipating spatial relationships~\cite{linnEmergenceCharacterizationSex2025}, we designed CareerPooler to leverage familiar physical dynamics: striking, collision, and trajectory naturally embody abstract career concepts such as effort, interconnection, and uncertainty, making the exploration process both intuitive and engaging.

In CareerPooler, users explore possible career development paths through a pool table-like interface: each ball on the table represents a career milestone or decision point; hovering over a ball reveals ambiguous ``hints'' that capture the partial visibility of future outcomes; and pocketing a ball symbolizes deliberate effort toward meaningful progress. The challenge of striking balls not only provides interactive difficulty but also reflects the reality of career exploration, where goals cannot simply be achieved by a single choice but require repeated attempts, occasional setbacks, and negotiation with external constraints. CareerPooler leverages large language models (LLMs) to generate contextual hints and narratives around each ball, translating user interactions on the table into career-relevant reflections. This design aligns with emerging paradigms in LLM-based interfaces, such as story generation through toy-based play~\cite{chungToytellerAIpoweredVisual2025} and world-building via ``dust and magnet'' metaphors~\cite{chungPatchviewLLMpoweredWorldbuilding2024}, extending these approaches by applying playful, spatially-grounded interaction to career exploration.

Based on prior research, we conducted a user study to investigate the potential of a pool-table metaphor for supporting career exploration in the context of users' self-selected career directions. In particular, we investigate three research questions:
\begin{itemize}
    \item \RQA
    \item \RQB
    \item \RQC
\end{itemize}

Through the investigation of these research questions, this paper offers the following contributions:
\begin{itemize}
    \item We introduce CareerPooler, a generative AI-powered career exploration system that employs a pool-table metaphor to transform abstract career trajectories into spatial, game-like interactions, capturing uncertainty, path dependency, and the need for repeated attempts.
    \item We design and implement a narrative event generation pipeline that integrates milestones, skills, and random events, balancing positive and negative outcomes while tailoring experiences to user background and goals.

    \item Through a within-subjects study with 24 participants, we provide empirical evidence that spatially-grounded metaphorical interactions can outperform conversational interfaces for career exploration in terms of engagement, satisfaction, and career clarity. We distill design implications for AI-assisted career exploration and, more broadly, for interactive systems that employ visual metaphors to support decision-making under uncertainty.
\end{itemize}

%% file: sections/2-relatedWork.tex
\section{Related Work}
Our study builds on three areas of prior research: (1) playfulness, uncertainty, and spatial interaction with generative systems, which establishes the methodological foundation for how game-based approaches and visual interfaces can render abstract processes tangible and engaging; (2) career exploration and challenges, focusing on factors that impact how individuals navigate dynamic opportunities and challenges to shape their professional trajectories; and (3) strategies and technologies for career exploration, encompassing traditional frameworks, HCI interventions, and emerging AI-based tools. These perspectives collectively inform our system design and investigation into how interactive systems can facilitate open-ended, meaningful career exploration.

\subsection{Playfulness, Uncertainty, and Spatial Interaction with Generative Systems}
\label{relatedplayful}

Uncertainty is widely recognized as a fundamental element that makes games engaging. Costikyan~\cite{costikyan2013uncertainty} argues that the struggle to master uncertainty is central to games' appeal, while To et al.~\cite{to2016integrating} connect this framework with curiosity theory, proposing that different types of uncertainty—including solver's uncertainty and hidden information—can motivate and accommodate players' varying curiosity levels. The application of playfulness to serious domains has demonstrated positive effects on learning outcomes and motivation across healthcare, education, and professional training~\cite{wouters2013meta, dichev2017gamifying}, with attention, active engagement, feedback, and consolidation identified as key design pillars~\cite{taly2017serious}. Notably, interactive fiction has been used to address sensitive topics by providing safe psychological distance while maintaining enough interactivity to shift users from passive absorbers to active participants~\cite{to2022interactive}. These findings suggest that playfulness can reduce psychological barriers and enable exploration of possibilities that might feel threatening in more formal contexts—a principle directly relevant to career exploration, where the stakes of ``trying out'' unfamiliar paths can feel prohibitively high.

Visual and spatial interaction techniques offer promising directions for making abstract AI-generated content more tangible. Prior studies have observed that AI outputs often appear overly comprehensive and verbose, hindering users' ability to extract salient information and exercise control. To address this, researchers have developed visual interaction techniques that structure and concretize generative processes. Sensecape~\cite{suhSensecapeEnablingMultilevel2023}, Luminate~\cite{suhLuminateStructuredGeneration2024}, and Graphologue~\cite{jiangGraphologueExploringLarge2023} employ graph- and tree-based layouts to reveal hierarchical relationships among generated elements. Intent-based user interfaces~\cite{dingIntentbasedUserInterfaces2024} treat user goals as first-class objects, with IntentPrism~\cite{wangIntentPrismHumanAIIntent2025} surfacing dynamic ``intent trees'' during web foraging to support progressive refinement.

These visual approaches build on established principles of interface metaphor design. Barr et al.~\cite{barr2002taxonomy} provide a systematic taxonomy of user-interface metaphors grounded in Lakoff and Johnson's framework, distinguishing between process metaphors that explain system functionality through real-world analogies and element metaphors that cue users into applicable interactions. A related strand of work examines visually interacting with generative systems for storytelling, where spatial and visual metaphors ground abstract narrative construction. TaleBrush~\cite{chungTaleBrushSketchingStories2022} demonstrates AI-assisted story sketching through freeform visual composition. Patchview~\cite{chungPatchviewLLMpoweredWorldbuilding2024} introduces the ``dust and magnet'' metaphor for world-building, while ToyTeller~\cite{chungToytellerAIpoweredVisual2025} generates emergent storylines through interactions with toy-like characters, highlighting the potential of embodied and playful visual metaphors for narrative generation. The integration of AI into playful experiences represents an emerging frontier: procedural content generation (PCG) has long been used in games to create varied experiences~\cite{shaker2016procedural}, and large language models expand possibilities for generating meaningful, personalized content~\cite{yannakakis2018artificial}. Research on human-AI co-creativity explores how AI can serve as a creative partner rather than merely a tool~\cite{maher2022designing, kantosalo2016modes}, while studies of playful interactions with ChatGPT have identified patterns such as ``contriving'' that position AI as a medium for creative exploration~\cite{yao2024interrogating}.

Our work extends these approaches by applying playful, spatially-grounded interaction to career exploration—a domain characterized by uncertainty, nonlinearity, and the need to balance aspiration with realism. We introduce a pool-table metaphor where generative AI serves as a narrative partner, generating contextual events, skills, and setbacks that emerge from the spatial dynamics of gameplay. The pool-table metaphor leverages familiar physical dynamics—striking, collision, trajectory—to embody abstract career concepts such as effort, interconnection, and uncertainty, making the exploration process both intuitive and engaging.

\begin{table}[!t]
\centering
\caption{Feature comparison across related systems that support AI-assisted career exploration. CareerPooler integrates features that are individually present in prior systems but have not been combined in a single career exploration tool. Note that the presence of a feature does not imply equivalent depth of implementation across systems.}
\label{tab:comparison-career}
\footnotesize
\renewcommand{\arraystretch}{1.1}
\begin{tabular}{@{}lcccccc@{}}
\toprule
\textbf{System} & \makecell{\textbf{Simulated}\\\textbf{Career Trajectories}} & \makecell{\textbf{Uncertainty}\\\textbf{Modeling}} & \makecell{\textbf{Gamified}\\\textbf{Exploration}} & \makecell{\textbf{Balanced}\\\textbf{Pos/Neg}} & \makecell{\textbf{Personalized}\\\textbf{Event}} & \makecell{\textbf{Reflective}\\\textbf{Learning}} \\
\midrule
Future You (FIE'25)~\cite{pataranutapornFutureYouConversation2024} & & & & & \cmark & \cmark \\
Letters from Future Self (CHI'25)~\cite{jeonLettersFutureSelf2025} & & & & & \cmark & \cmark \\
CareerSim (CHI EA'24)~\cite{du2024careersim} & \cmark & & \cmark & & \cmark & \\
VizCV (arXiv'25)~\cite{lazrik2025} & & & & & & \cmark \\
BlindTokers (CHI'26)~\cite{lyu2026mybrotherschoolprincipal} & & \cmark & & \cmark & & \cmark \\
Junior to Senior (CHI'26)~\cite{feng2026} & & \cmark & & \cmark & & \cmark \\
\textbf{CareerPooler} & \cmark & \cmark & \cmark & \cmark & \cmark & \cmark \\
\bottomrule
\end{tabular}

\vspace{1em}

\caption{Feature comparison across related systems that support interactive narrative generation. These features represent distinct design dimensions; their co-presence in CareerPooler reflects our specific application to career exploration, where uncertainty, partial information, and progressive disclosure are domain requirements rather than optional enhancements.}
\label{tab:comparison-narrative}
\renewcommand{\arraystretch}{1.1}
\begin{tabular}{@{}lcccccc@{}}
\toprule
\textbf{System} & \makecell{\textbf{Spatial}\\\textbf{Metaphor}} & \makecell{\textbf{Non-Linear}\\\textbf{Narrative}} & \makecell{\textbf{Partial}\\\textbf{Information Hints}} & \makecell{\textbf{Embodied/Playful}\\\textbf{Interaction}} & \makecell{\textbf{Emergent}\\\textbf{Storytelling}} & \makecell{\textbf{Progressive}\\\textbf{Disclosure}} \\
\midrule
Sensecape (UIST'23)~\cite{suhSensecapeEnablingMultilevel2023} & \cmark & & & & & \cmark \\
Graphologue (UIST'23)~\cite{jiangGraphologueExploringLarge2023} & \cmark & & & & & \\
Luminate (CHI'24)~\cite{suhLuminateStructuredGeneration2024} & \cmark & & & & \cmark & \cmark \\
PatchView (UIST'24)~\cite{chungPatchviewLLMpoweredWorldbuilding2024} & \cmark & \cmark & & \cmark & \cmark & \\
Toyteller (CHI'25)~\cite{chungToytellerAIpoweredVisual2025} & & \cmark & & \cmark & \cmark & \\
Sketchar (CHI PLAY'24)~\cite{sketchar2024} & & \cmark & & \cmark & \cmark & \cmark \\
\textbf{CareerPooler} & \cmark & \cmark & \cmark & \cmark & \cmark & \cmark \\
\bottomrule
\end{tabular}
\end{table}

\subsection{Career Exploration and Challenges in the AI Era}
Career exploration involves eliciting and sustaining personal interests within dynamic environments, gathering information about oneself and occupational landscapes, and constructing self-identity through engagement with opportunities and challenges~\cite{stumpf1983development, zikic2006job}. 
As a lifelong process, it re-emerges during transitions, enabling individuals to adapt career roles to evolving personal goals and external conditions~\cite{super1980life, jiang2019career}.
Its effectiveness depends on both external factors (e.g., economic trends, technological advancements, labor policies) and internal barriers that vary by life stage~\cite{jiang2019career}. For instance, emerging adults (ages 18-25) often struggle with limited financial resources, sparse professional networks, skill deficits, or are unclear about self-identity~\cite{bloomer2021trying, dhakal2018inclusion}, while experienced workers may encounter ``career stuckness''~\cite{treebak2022towards}, fear of losing established status, and reluctance to seek support~\cite{peake2012chaotic, masdonati2022involuntary}.

Recent advances in generative AI introduce both new anxieties and new possibilities. On one hand, generative AI tools heighten the uncertainty in industry and workplace: recent studies document a decline of over 13\% in entry-level job postings in fields such as software development and customer service since the rapid adoption of AI in 2022, reducing traditional pathways for gaining foundational experience~\cite{brynjolfsson2025canaries}. Integrating AI into daily workflows also generates stress and necessitates renewed exploration of evolving roles. Research shows that awareness of STARA (Smart Technology, Artificial Intelligence, Robotics, and Algorithms) technological replacement risk is linked to increased job-related anxiety and diminished well-being~\cite{jin2024work}.
On the other hand, AI-powered tools offer accessible and cost-effective forms of support. Large language models (LLMs) can provide personalized guidance through skills mapping, résumé optimization, and simulated interview practice, enhancing readiness and confidence~\cite{gomez2025virtual}.
Generative AI thus acts as a double-edged sword: while reshaping labor markets in ways that compel individuals to reconsider and reconstruct their career trajectories, it simultaneously offers unprecedented support for career exploration. Yet, its long-term costs and benefits for individual career explorers remain underexplored. This work investigates how AI-generated content can be designed to acknowledge career uncertainty rather than obscure it, helping users engage with both opportunities and setbacks in a balanced manner. These challenges—particularly the heightened uncertainty and the inadequacy of static, idealized guidance—directly motivate CareerPooler's design: a system that surfaces both positive and negative outcomes, models path dependency, and uses spatial interaction to make uncertainty tangible rather than hidden.

\subsection{Strategies and Technologies Supporting Career Exploration}
Traditional strategies supporting career exploration often emphasize the alignment between personal attributes and professional opportunities. Self-assessment tools like Holland's RIASEC model and Myers–Briggs Type Indicator (MBTI) map personality traits to careers. However, these instruments often reduce complex identities to static categories and show context-dependent validity which limit their reliability across diverse populations and life stages~\cite{gardner1996using}. 
Career counseling provides more tailored support by combining personal aspirations with labor market realities, while networking, internships, and mentoring provide valuable exposure and skill development~\cite{maree2020innovating}. However, these approaches require substantial personal and financial investment~\cite{raposa2018economic}, which places individuals without such resources or those who simply want to explore at a distinct disadvantage.

Human-computer interaction (HCI) tools address these gaps by leveraging interactive technologies. My Automated Conversation coacH (MACH) delivers real-time feedback to enhance communication skills, simulating professional interactions~\cite{hoque2013mach}. Virtual career fairs and online platforms like LinkedIn and Indeed connect job seekers with employers, broadening access to opportunities~\cite{sen2022virtual, ladkin2016online, linkedin}. Digital portfolios, such as Mahara and PebblePad, enable users to document skills, reflect on experiences, and present achievements in dynamic, multimedia formats~\cite{mahara, pebblepad}. Large language models (LLMs) introduce novel affordances: AI-powered résumé feedback systems provide tailored suggestions~\cite{awasthi2025resumegenai}, while conversational agents like Steve support interview preparation, skills assessments, and upskilling pathways~\cite{renji2025steve}. SkillsIdentifier~\cite{dillahunt2021skillsidentifier} helps underrepresented job seekers identify and articulate their skill sets, demonstrating that such tools can foster career identity and self-efficacy. Reflective practices, such as AI-driven ``letters to future self'' or guided conversations, reduce career anxiety and boost motivation, particularly for emerging adults~\cite{pataranutapornFutureYouConversation2024, jeonLettersFutureSelf2025}.

In career development specifically, existing playful interventions have primarily focused on gamified elements like badges and points, or simple career exploration games for younger students~\cite{euroguidance2024, brandenburger2023gamified}. More recent work has begun exploring gamified career simulations using AI, such as CareerSim~\cite{du2024careersim}, which employs role-playing game mechanics to encourage reflection on career decisions. However, these approaches often rely on traditional RPG interfaces and have not fully explored how spatial metaphors might transform the career exploration experience.

Despite these advances, significant limitations persist. Many tools cater to affluent job seekers~\cite{dillahunt2021skillsidentifier} and require rich input such as detailed resumes, skill inventories, or step-by-step plans, creating barriers for people who may not yet be ready to share such information or lack the experience to articulate it clearly. Moreover, many tools emphasize discrete tasks such as resume refinement, skill-gap analysis, or interview rehearsal, often neglecting the exploratory, nonlinear aspects of career development. Systems frequently frame careers in static or overly prescriptive ways, reducing identity development to fixed checkpoints rather than ongoing journeys. At the same time, interactions with these systems are often serious and transactional, limiting curiosity and engagement, and the guidance produced is frequently vague or overly idealized.

CareerPooler addresses these specific gaps in three ways: (1) it lowers the barrier to entry by requiring only a brief self-description and career aspiration rather than detailed resumes or skill inventories; (2) it transforms the exploration from a serious, transactional process into a playful, spatially-grounded interaction that sustains curiosity and engagement; and (3) it generates balanced career narratives that include setbacks and uncertainty rather than only idealized outcomes, combining the interaction paradigm introduced in Section~\ref{relatedplayful} with the career exploration context discussed above. Table~\ref{tab:comparison-career} and Table~\ref{tab:comparison-narrative} summarize how Careerpooler compares with related work across representative features.

%% file: sections/3-systemDesign.tex
\section{System Design}

The system is designed to support career development simulation through spatial interaction and event generation. The overarching objective is to provide an engaging and analytically meaningful environment for exploring potential career trajectories. To achieve this, we formulated three design goals, each of which is operationalized in the subsequent system implementation.
 \begin{figure}
 \centering
 \includegraphics[width=1\textwidth]{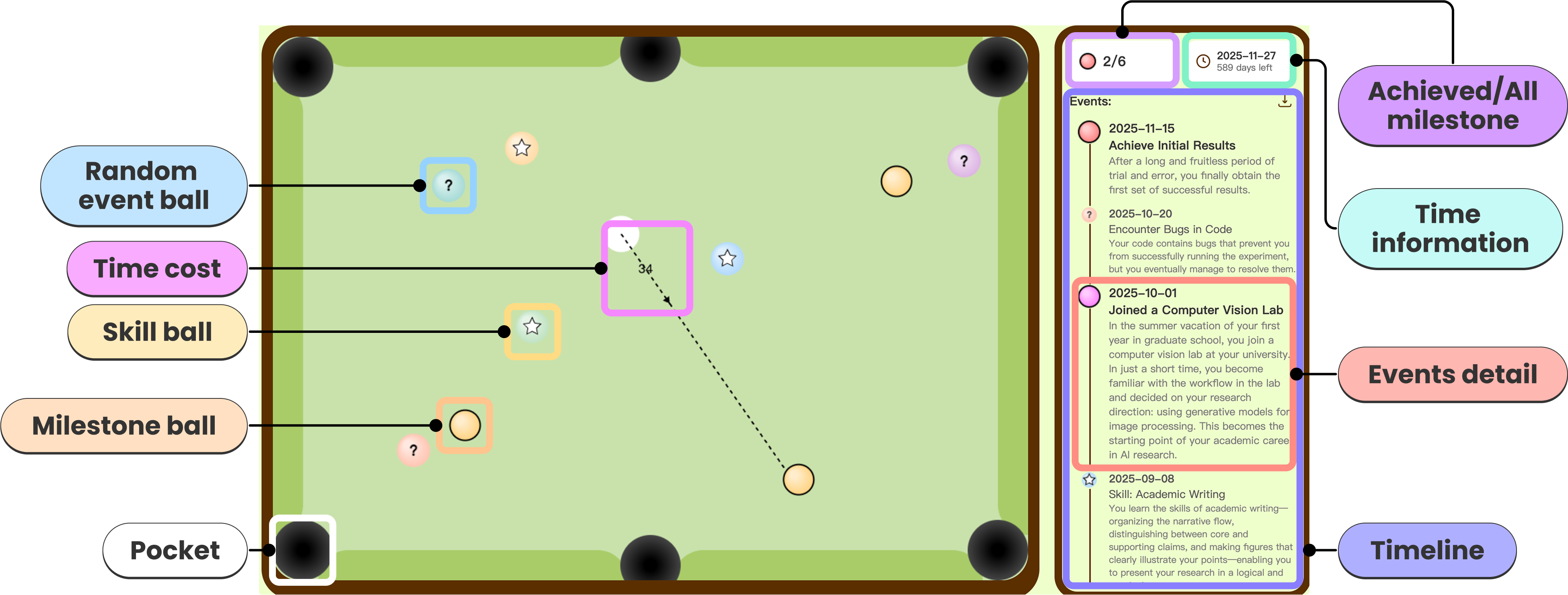}
   \caption{The main interface layout and core functional modules of CareerPooler. Including Game Mechanism Elements (left): Random event ball, Time cost line, Skill ball, Milestone Ball, and Pocket. And Information Display Modules (right): Achieved/All milestone, Time Information, Timeline and Events detail within the timeline.}
   \Description{The main interface layout and core functional modules of CareerPooler.}
   \label{fig:pooltable}
 \end{figure}

\subsection{Design Goals}

Our design goals are informed by three strands of prior research. First, work on serious games and gamification has shown that playful interaction increases engagement and motivation in learning contexts~\cite{wouters2013meta, taly2017serious}, while interactive fiction provides psychological safety for exploring sensitive topics~\cite{to2022interactive}—suggesting that a game-based approach can lower the emotional barriers of career exploration. Second, Costikyan~\cite{costikyan2013uncertainty} and To et al.~\cite{to2016integrating} establish that uncertainty is central to what makes games compelling; spatial and visual metaphors further help users navigate abstract AI-generated content by making structure and relationships tangible~\cite{suhSensecapeEnablingMultilevel2023, barr2002taxonomy}. Third, career development research highlights that emerging adults face heightened uncertainty due to AI-driven labor market shifts~\cite{brynjolfsson2025canaries, shaoFutureWorkAI2025}, yet existing tools often produce overly idealized or comprehensive guidance that neglects the nonlinear, setback-laden reality of career trajectories~\cite{du2024careersim}. These insights collectively motivate the following three design goals:

\begin{enumerate}
    \item \textbf{DG1: Design an engaging and satisfying career exploration system for early career participant.}  
    The system should provide an experience that is engaging and satisfying, encouraging users—particularly those at early career stages—to actively explore potential career trajectories in an adaptive and risk-free environment.  

    \item \textbf{DG2: Simulate real-world uncertainty through spatial interaction and partial information.} The system should incorporate uncertainty into the spatial interaction process, with outcomes remaining partially unpredictable and informed only by limited hints. This approximates the ambiguity of real-world career development and supports decision-making under incomplete information.
 
    \item \textbf{DG3: Ensure realistic and personalized experience generation to improve clarity in career exploration.}  
    The system should generate events that are contextually realistic and tailored to user background, goals, and prior choices, with a balanced distribution of successful, unsuccessful, and neutral outcomes. Such generation improves the clarity of simulated career exploration and enhances its educational value.  
\end{enumerate}

\subsection{System Implementation}

The system implementation follows the three design goals, with specific components addressing each.


\subsubsection{Interaction Design (DG1, DG2)}

The interaction layer employs a pool spatial metaphor (DG1, DG2), grounded in the observation that pool inherently combines skill-based action with chance-based outcomes~\cite{caillois1961man}—the same interplay of deliberate effort and unpredictability that characterizes real-world career development. In the language of procedural rhetoric~\cite{bogost2007persuasive}, the pool table does not merely represent career concepts but enacts them: the drag distance of the cue ball corresponds to the number of days invested, linking striking force to time commitment; collisions and rebounds introduce uncertainty that mirrors unexpected outcomes in career trajectories; ball trajectories embody path dependency, as each shot's outcome constrains subsequent options; pockets represent milestone achievement, requiring both aim and effort; and table boundaries act as real-world constraints, with rebounds mirroring how external limitations redirect one's course. Optimized collision detection and high-frequency rendering ensure consistent and responsive interaction. To engage users and provide a satisfying experience (DG1), the system combines visual feedback, progressive challenge, and cumulative outcomes to sustain user attention and motivate exploration. Event balls are semantically differentiated into four categories (DG2):

\begin{itemize}
    \item \textbf{Milestone balls:} Represent major career stages, such as receiving a PhD offer.
    \item \textbf{Skill balls:} Represent potential skill acquisitions, such as learning to write a cold email.
    \item \textbf{Random event balls:} Represent unforeseen events with positive, neutral, or negative outcomes, such as a funding cut.
    \item \textbf{Decision event balls:} A subset of random events requiring binary decisions, such as whether to follow a mentor who changes institutions.
\end{itemize}

The interface (see Figure \ref{fig:pooltable}) consists of a dual-panel layout (DG2): the left panel contains the pool table where actions occur, while the right panel displays a timeline and event log for reflection. To approximate the limited foresight of real-world planning, the system integrates an information hints mechanism (DG2): hovering over skill or random event balls reveals short abstract hints about potential outcomes without disclosing their sentiment polarity or consequences. This preserves uncertainty while enabling bounded foresight, thereby supporting engagement (DG1).

The simulation ends when either six milestones have been collected or 730 days have elapsed. The two-year (730-day) timeframe reflects the short-cycle, iterative nature of career exploration during emerging adulthood, a stage characterized by frequent reassessment and identity experimentation~\cite{super1980life, arnett2000emerging}. Meaningful career transitions—such as completing a degree, changing roles, or acquiring key competencies—typically unfold within one- to two-year windows for this population~\cite{murphy2010realistic}. The six-milestone cap corresponds to approximately one major career event every four months, mirroring the realistic pace of early-career developments such as passing a qualifying exam, publishing a paper, or receiving a job offer~\cite{savickas2012life}. Upon termination, the system generates a career journey report that summarizes collected milestones, skills, random events, and provides analytical reflections.

\subsubsection{Event Generation Pipeline (DG3)}

\begin{figure}
 \centering
 \includegraphics[width=1\textwidth]{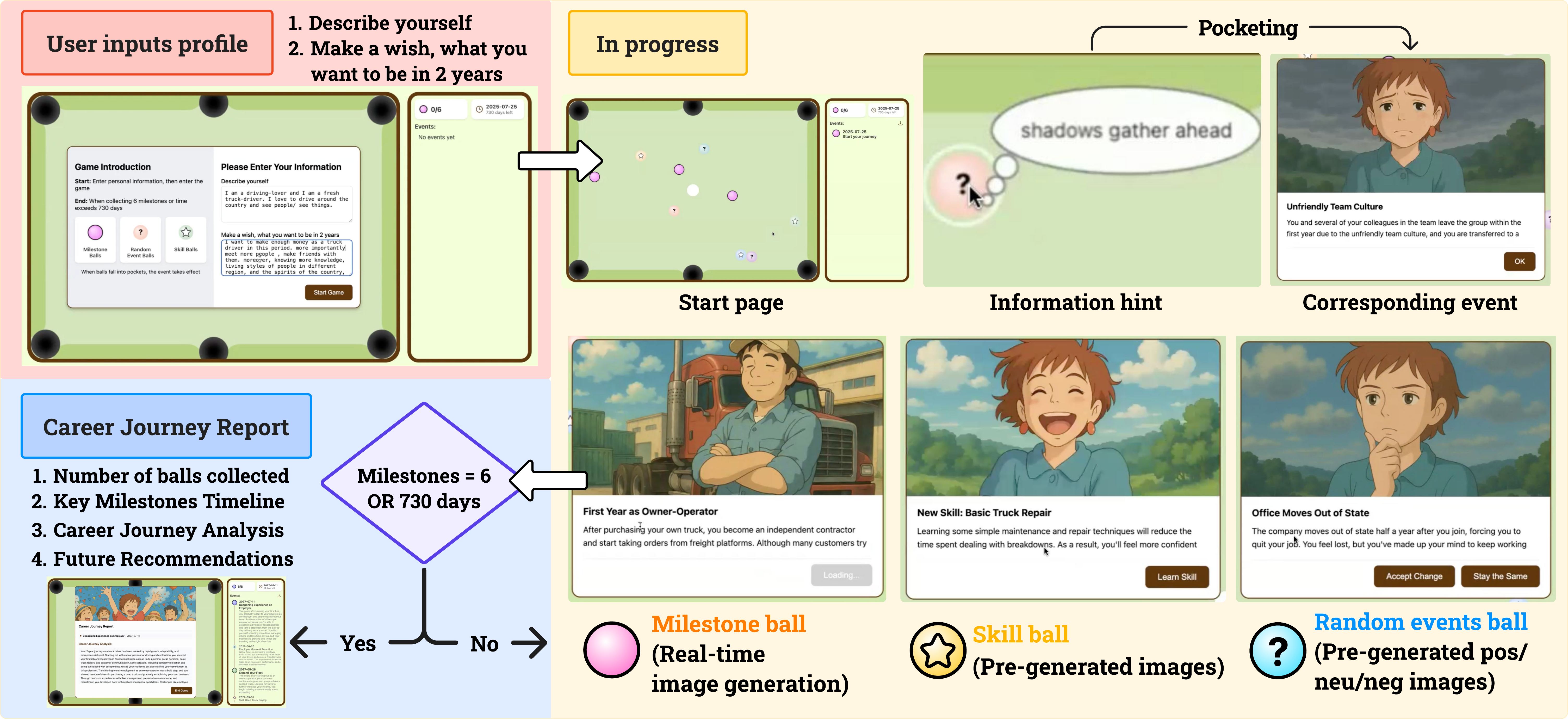}
   \caption{Workflow of System Interaction and Event Generation: After providing personal background and a two-year career goal, users enter the interactive interface. Users can hover over skill or random event balls to preview information hints, or pocket balls to collect events. Collected events include: skill balls, milestone balls, and random event balls. The simulation ends when either six milestones are collected or 730 days have elapsed, at which point a career journey report is generated.}
   \Description{Workflow of system interaction and event generation.}
   \label{fig:workflow}
 \end{figure}

The event generation pipeline produces structured and coherent events using large language and vision models, as shown in figure\ref{fig:workflow}. Each round generates one milestone, three random events, and three skill events. Random and skill events are paired with abstract hints consistent with the information hints mechanism. Generated events follow an output format (``title: \$\{event title\} | \$\{event content\}  \$\{image label\}'') for parsing.  

Two constraints guide generation: (1) all events must fit within the two-year simulation window, and (2) events must maintain logical dependence on previously triggered events. For visual content, a hybrid strategy is used: milestone images are generated in real time (see Appendix~\ref{appendix:b}), while random and skill events are pre-generated by sentiment category to minimize latency. Asynchronous processing ensures gameplay is not blocked by model inference. Since large language models may overrepresent successful trajectories due to survivorship bias in their pretraining corpora, the system employs mitigation strategies. In particular, fine-tuning examples and prompts (see Appendix~\ref{appendix:a}) were designed to explicitly incorporate unsuccessful and neutral narratives, thereby encouraging more balanced and realistic event generation.  

%% file: sections/4-studyDesign.tex
\section{Study Design}

To evaluate CareerPooler's effectiveness in facilitating career exploration and planning, we conducted a within-subjects comparative study examining how the system's spatial interface and fine-tuned career event generation influence users' career exploration experiences compared to a traditional linear chat interface. Table~\ref{tab:demographics} provides demographic information. 24 participants in our study represented diverse career stages and professional backgrounds. The sample included 3 working professionals (1 UX designer, 1 firmware engineer, and 1 testing engineer), 20 students across different educational levels (3 undergraduate students, 11 Master's students, 5 PhD students, 1 postdoctoral student), and 1 participant currently unemployed.

\begin{table}[h]
\centering
\caption{Participant demographics (N = 24).}
\label{tab:demographics}
\resizebox{\textwidth}{!}{%
\begin{tabular}{|l|c|c|l|l|l|c|c|l|l|}
\hline
\textbf{ID} & \textbf{Age} & \textbf{Gender} & \textbf{Current occupation} & \textbf{Specialization} & \textbf{ID} & \textbf{Age} & \textbf{Gender} & \textbf{Current occupation} & \textbf{Specialization} \\
\hline
P1 & 27 & F & UX Designer & UX Research & P13 & 24 & F & PhD student & AR/VR \\
P2 & 20 & M & Undergraduate student & NLP & P14 & 24 & M & Master student & Robotics \\
P3 & 24 & F & Master student & Human-AI Interaction & P15 & 25 & F & Master student & UI/UX \\
P4 & 24 & F & Master student & Data Visualization & P16 & 24 & F & Master student & Illustration \\
P5 & 23 & M & Undergraduate student & Social Computing & P17 & 24 & M & Testing engineer & Industrial Engineering \\
P6 & 32 & F & PhD student & Accessibility and Inclusive Design & P18 & 28 & F & Firmware engineer & Mechanical Engineering \\
P7 & 28 & F & Unemployed & Marketing & P19 & 23 & M & Master student & Software Engineering \\
P8 & 23 & F & Master student & Electronic Music & P20 & 24 & F & Master student & AI \\
P9 & 25 & M & Master student & Psychology & P21 & 25 & M & Master student & Social Computing \\
P10 & 26 & F & PhD student & Conversational Agent & P22 & 32 & M & PhD student & UX Researcher \\
P11 & 26 & M & Postdoc & LLM & P23 & 21 & F & Undergraduate student & NLP \\
P12 & 26 & M & Master student & UI & P24 & 25 & F & PhD student & Human-Robot Interaction \\
\hline
\end{tabular}%
}
\end{table}

\begin{table}[t]
\centering
\caption{Research questions, survey questions (SQs), and interview questions (IQs). The study employs a mixed-methods design, combining quantitative 5-point Likert scale surveys with qualitative interviews to evaluate CareerPooler relative to the baseline system.}
\label{tab:interviewquestion}
\footnotesize
\renewcommand{\arraystretch}{1.3}
\begin{tabular}{@{}p{1.8cm}p{3.5cm}p{5.2cm}@{}}
\toprule
\textbf{Dimension} & \textbf{Survey Question} & \textbf{Interview Question} \\
\midrule
\rowcolor{red!15}
\multicolumn{3}{@{}l}{\textbf{RQ1: How does a spatially gamified AI career simulation influence user experience compared to AI chatbots?}} \\
\midrule
Engagement & SQ1: Interacting with this system gave me new insights or unexpected realizations about my career path. & IQ1: After using each system, did you gain any new insights or unexpected realizations about your career path? If so, could you share an example? \\
Information Gain & SQ2: Interacting with this system kept me engaged throughout the experience. & IQ2: In what ways did each system influence your engagement in the experience? \\
Authenticity & SQ3: The career events generated by this system felt realistic and believable. & IQ3: How believable or authentic did you find the career events generated by each system? Can you recall a specific event that stood out to you as especially realistic or unrealistic? \\
Satisfaction & SQ4: Overall, I was satisfied with my experience using this system. & IQ4: Looking at your overall experience, what specific features or moments contributed most to your satisfaction or dissatisfaction? What improvements would you suggest for both? \\
\midrule
\rowcolor{yellow!25}
\multicolumn{3}{@{}l}{\textbf{RQ2: How does such a system influence career exploration outcomes compared to AI chatbots?}} \\
\midrule
Career Clarity & SQ5: After this experience, I have a clear direction for my career. & IQ5: After planning with both tools, which one left you with clearer career goals, and why? \\
Step-wise Understanding & SQ6: After this experience, I understand the key steps I need to take to achieve my career goals. & IQ6: Which tool better helped you understand the key steps needed to achieve your career goals, and why? \\
Career Confidence & SQ7: After this experience, I am confident in my ability to achieve the goals I set for myself. & IQ7: Which tool made you feel more confident about carrying out your plan, and what features contributed to that confidence? \\
\midrule
\rowcolor{blue!10}
\multicolumn{3}{@{}l}{\textbf{RQ3: How do individuals perceive AI's impact on careers, and how do they prepare for it?}} \\
\midrule
AI Awareness & & IQ8: Have people around you experienced a career change due to AI? If so, in what ways were they affected? Can you provide an example? \\
AI Strategies & & IQ9: How do you view AI's impact on the job market, and how do you take those factors into account when making your career plan? \\
\bottomrule
\end{tabular}
\end{table}

\subsection{Participants}

We recruited 24 participants using a mixed-method approach combining online recruitment, offline methods, and snowball sampling. Initial participants were identified through various online channels and offline methods, who then referred additional eligible participants from their networks. Our target population consisted of students and early-career professionals experiencing uncertainty about their future career trajectories, as this demographic represents primary users of career exploration tools. Participants ranged in age from 20 to 32 years (M = 25.13, SD = 2.83). 

\subsection{Task Design}

We employed a within-subjects design with counterbalanced ordering to control for potential learning effects. Each participant experienced both conditions:

\textbf{Condition A - CareerPooler}: Participants used the CareerPooler system, which features:
\begin{itemize}
    \item A spatial pool-style interface for exploring career trajectories;
    \item Fine-tuned LLM based on career development data;
    \item Visual representation of career paths as interconnected pools;
    \item Interactive exploration through the metaphor of ``banking shots'' between career states.
\end{itemize}

\textbf{Condition B - ChatGPT (baseline)}: Participants used a linear chat interface powered by GPT-4o. We provided participants with semi-structured prompts for career exploration with ChatGPT (see Appendix~\ref{appendix:c}). We selected ChatGPT as our baseline because it is the most widely adopted AI assistant among our target population—students and early-career professionals—and no existing AI tools specifically support spatially gamified career exploration. ChatGPT thus represents current practice in AI-assisted career guidance, where users typically seek general AI advice through conversational interfaces without spatial or game-based interaction~\cite{pataranutapornFutureYouConversation2024, du2024careersim}. The order of conditions was counterbalanced: half of the participants (n = 12) experienced CareerPooler first, while the other half (n = 12) began with the baseline condition.

For each condition, participants were instructed to explore potential career trajectories over a two-year timeframe. We chose a two-year horizon based on career development literature suggesting this timeframe balances near-term actionability with sufficient scope for meaningful career transitions. Participants were required to explore at least two distinct career paths per system but were encouraged to explore additional trajectories if desired. This design decision allowed us to observe both breadth (multiple path exploration) and depth (detailed exploration within paths) of career exploration behavior. No upper limit was imposed on the number of explorations to avoid constraining natural exploration patterns.

\subsection{Study Procedure}

\begin{figure}
\centering
\includegraphics[width=1\textwidth]{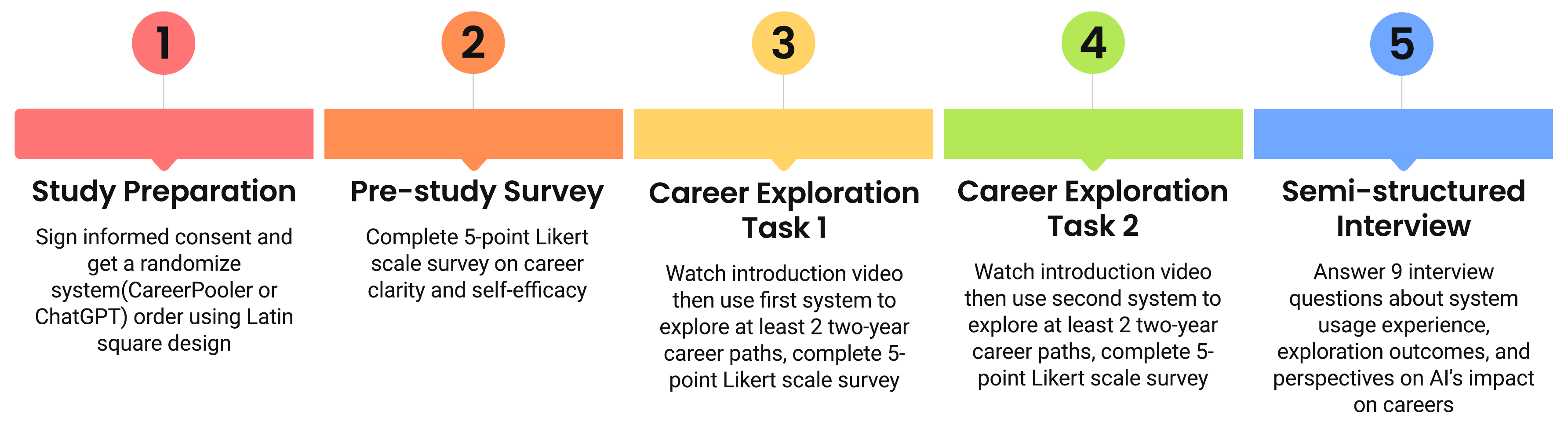}
  \caption{Study workflow: participants completed five phases: (1) study preparation with randomized system assignment; (2) pre-study survey on career clarity; (3-4) two career exploration tasks using CareerPooler and ChatGPT in counterbalanced order, each with post-surveys; and (5) semi-structured interview on experiences and AI's career impact. Total session time averaged 105 minutes.}
  \Description{StudyWorkflow}
  \label{fig:StudyWorkflow}
\end{figure}

The study procedure is shown in Figure \ref{fig:StudyWorkflow}. Participants first reviewed and signed informed consent forms before completing a pre-study assessment measuring baseline career clarity and self-efficacy using adapted items from the Career Decision Self-Efficacy Scale and Career Clarity Scale on 5-point Likert scales (e.g., ``I have a clear direction for my career", ``I understand the key steps I need to take to achieve my career goals", ``I am confident in my ability to achieve the goals I set for myself"). In the first round, participants used their assigned first system (CareerPooler or Baseline) to explore career trajectories following a brief tutorial. Immediately after, they completed a mid-study assessment measuring both user experience dimensions (Engagement, Information Gain, Perceived Authenticity, Overall Satisfaction) and career exploration outcomes (Career Clarity, Actionable Understanding, Implementation Confidence). Participants then repeated the career exploration task using the alternate system in second round, followed by the same assessment to enable within-subjects comparison. The session concluded with semi-structured interviews. The mapping between research questions, Likert scale statements and interview questions are shown in Table~\ref{tab:interviewquestion}. Study durations varied depending on participants' engagement with the CareerPooler system, as no time limit was imposed. The average study duration was 107.39 minutes ($\min=44$, $\max=404$).

\subsection{Procedural Ethical Considerations}

Given the nature of our study, which relates to participants’ personal development and career exploration, we carefully considered ethical aspects in the design and procedure. All sessions were conducted in a supportive and user-friendly manner. The procedure is designed to provide participants with useful and constructive career information through the experiment, so that their engagement could also yield personal benefits. Participants were encouraged to freely express their own preferences and could stop or withdraw at any point without consequence. To minimize potential risks, no sensitive personal identifiers were collected, and all responses were anonymized. Our goal was to ensure that the study not only generated valuable research insights but also respected participants’ autonomy, privacy, and well-being during the exploration process. We further elaborate on these ethical considerations in Section~\ref{sec:ethical_considerations}.

\subsection{Data Collection and Analysis}

We collected both user study interaction data and interview data. The study interaction data included screen recordings capturing participants' career exploration processes in both CareerPooler and Baseline systems, quantitative survey responses collected at three time points (pre-study, mid-study, and post-study), and the interaction data and content data with the CareerPooler system. All screen and audio recordings were obtained with participants' informed consent. For the interview data, we conducted and recorded semi-structured interviews following our protocol, which were automatically transcribed using Zoom's transcription service and manually corrected for accuracy. We used both quantitative and qualitative methods to examine the user study data:

\begin{itemize}
    \item \textbf{User interactions}: We examined interaction data including the number of shots taken, number of balls pocketed, and the detailed breakdown of different types of balls. This analysis helped us understand how participants engaged with the CareerPooler system.
    
    \item \textbf{Likert survey responses}: We analyzed the Likert-scale responses collected across three time points. This involved computing descriptive statistics for each condition, conducting paired comparisons between CareerPooler and Baseline systems, and examining changes from pre-study baseline to post-condition assessments. Effect sizes were calculated to assess the practical significance of observed differences.
    
    \item \textbf{Semi-structured interviews}: We conducted a thematic analysis of the interview responses based on Braun and Clarke's six-phase framework \cite{braunUsingThematicAnalysis2006}, with assistance of large language models. As a pre-screening step, Claude-Opus-4.1 was employed to generate a preliminary set of 20 potential themes derived from the interview transcripts, structured in relation to the guiding interview questions. Subsequently, the first and last authors collaboratively refined and organized these themes in alignment with the overarching research questions and sub-questions, reaching consensus through iterative discussion. Representative quotations were selected to illustrate each theme and are reported in the paper. Analytic memos were kept throughout the process to ensure transparency and rigor.
\end{itemize}

%% file: sections/5-results.tex
\section{Results}

\begin{figure}
\centering
\includegraphics[width=1\textwidth]{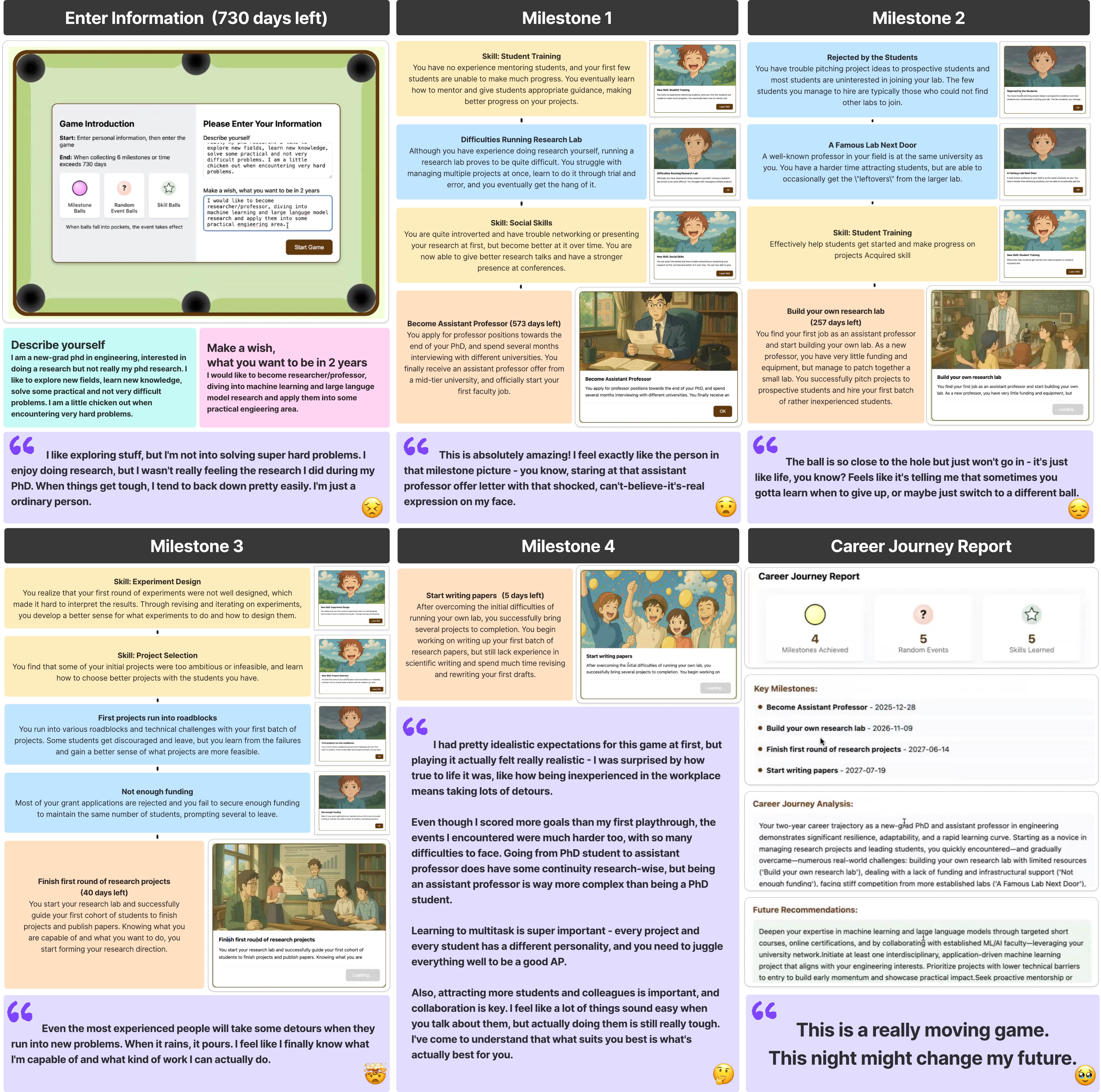}
  \caption{User journey of P11 over a 730-day simulation in CareerPooler. The participant began by providing a self-description and a career aspiration (“to become a researcher/professor specializing in machine learning and large language models”). In the early phase, P11 developed mentoring and social skills while struggling with initial student training and lab management. Midway, P11 obtained an assistant professor position, established a small research lab, and faced challenges such as attracting students, competing with a renowned neighboring lab, and securing limited funding. In subsequent milestones, P11 guided the first student cohort to complete projects, learned to design experiments more effectively, and refined project selection strategies despite technical setbacks and student attrition. In the final stage, P11 began preparing and revising research papers, marking the transition toward scholarly productivity. The journey concluded with a personalized career report summarizing milestones achieved, random events encountered, and skills acquired, reflecting both the accomplishments and difficulties of early academic career development.}

  \Description{Career Pooler.}
  \label{fig:userjourney}
\end{figure}

Through an analysis of quantitative measures and qualitative feedback from 24 participants in a comparative study, we examined three research questions to understand how CareerPooler's gamified approach influences career exploration compared to ChatGPT as AI chatbot interface. The results show that CareerPooler significantly outperforms ChatGPT across two dimensions of user experience: engagement ($p<<0.01$) and satisfaction ($p<<0.01$), while also enhancing participants' career clarity ($p=0.01$). The qualitative findings further contextualize these results by identifying the underlying reasons and contributing factors behind the statistical outcomes for RQ1 and RQ2, while also revealing the diverse strategies participants employed to navigate AI-mediated career exploration environments for RQ3.

\subsection{User Journey}

To illustrate how users interact with the system, Figure \ref{fig:userjourney} illustrates the simulated career trajectory of P11, a recent PhD graduate who aspired to become an assistant professor. The simulation began with an initial self-introduction and a career aspiration statement (“to become a researcher/professor specializing in machine learning and large language models”). Over the course of the 730-day simulation, P11 engaged with a series of milestones, skills, and random events that collectively represented both progress and setbacks. The first milestone centered on developing student mentoring and social skills, reflecting the challenges of guiding inexperienced students and presenting research within academic networks. The second milestone focused on obtaining a faculty position and establishing a new research lab. Despite eventually securing an assistant professor offer, P11 faced difficulties in attracting students, competing against a neighboring renowned lab, and coping with limited resources. These experiences underscored the competitive and resource-constrained environment of early faculty life.  

Subsequent milestones further emphasized the iterative and uncertain nature of academic research. In building the lab, P11 encountered difficulties in pitching projects to students, managing underfunded projects, and designing effective experiments. These challenges required developing competencies in experiment design, project selection, and lab management, while also dealing with high student turnover due to funding shortages. By the third milestone, P11 began to consolidate their research direction through guiding the completion of initial projects, although this process involved repeated trial-and-error and setbacks.

The culmination of the simulation was a personalized “Career Journey Report” summarizing the acquired skills, encountered random events, and achieved milestones. Throughout the journey, P11 reflected on the metaphorical resonance between the pool-inspired gameplay and academic life. For example, the difficulty of pocketing balls was compared to detours and elusive goals in a career, while the closeness of a ball to a pocket that still refuses to fall was likened to seemingly attainable but persistently unachieved aspirations. These reflections highlight how the embodied and spatial nature of the simulation rendered abstract career dynamics tangible. P11 described the experience as “really moving”, expressing that the game might influence future career decisions by reinforcing the realization that “what suits you best is actually the best.”

 \begin{figure}
 \centering
 \includegraphics[width=1\textwidth]{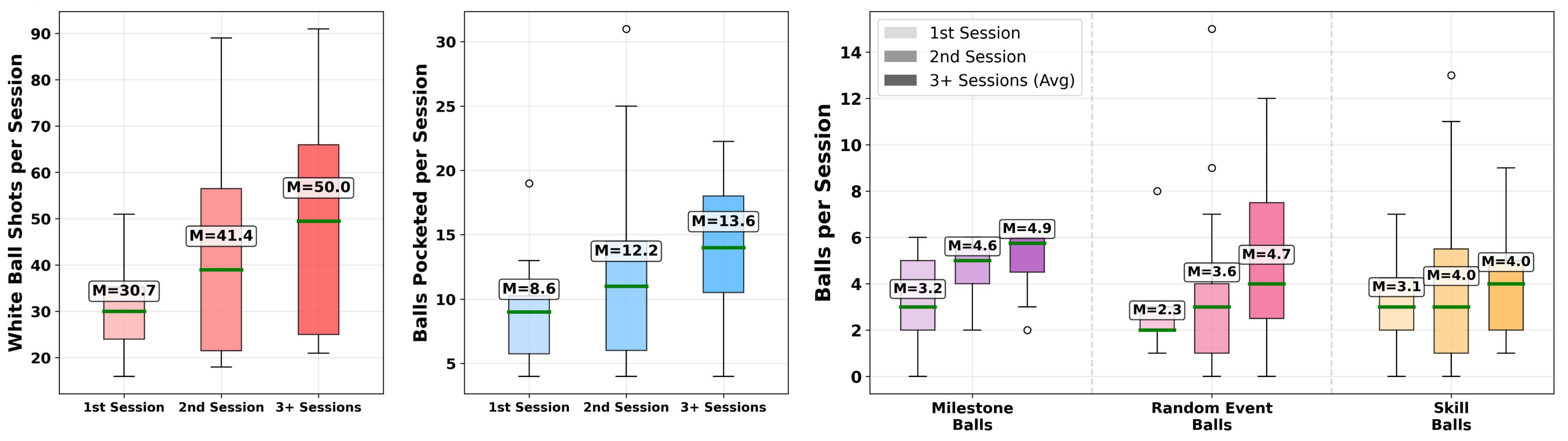}
   \caption{This figure presents the performance progression across different ball types and gaming sessions. For White Ball Shots per Session, participants showed improvement from the first session (M=30.7) to the most experienced players (3+ sessions, M=50.0). Balls Pocketed per Session increased from M=8.6 in the first session to M=13.6 for participants with three or more sessions. The ball type analysis shows progression across all three categories. Milestone Balls increased from M=3.2 (1st session) to M=4.9 (3+ sessions). Random Event Balls demonstrated improvement from M=2.3 to M=4.7 across sessions. Skill Balls showed growth from M=3.1 to M=4.0 across multiple gaming sessions.
   }
   \label{fig:ballanalysis}
 \end{figure}
 
\subsection{RQ1: CareerPooler significantly enhances user experience in engagement and overall satisfaction}

\begin{table}[h]
\centering
\caption{RQ1: User experience rating comparison between ChatGPT and CareerPooler. CareerPooler demonstrates significant difference in engagement and satisfaction, while information gain and perceived authenticity shows no significant difference.}
\label{tab:user-experience}
\begin{tabular}{lcccc}
\hline
\multirow{2}{*}{\textbf{Dimension}} & \textbf{ChatGPT} & \textbf{CareerPooler} & \multirow{2}{*}{\textbf{W}} & \multirow{2}{*}{\textbf{p-value}} \\
 & $(M \pm SD)$ & $(M \pm SD)$ & & \\
\hline
Engagement & $3.67 \pm 1.01$ & $4.54 \pm 0.59$ & 21.0 & $<< 0.01$** \\
Information Gain & $3.71 \pm 0.96$ & $4.17 \pm 0.87$ & 29.0 & $0.07$ \\
Perceived Authenticity & $3.75 \pm 1.03$ & $4.04 \pm 1.08$ & 39.5 & 0.22 \\
Overall Satisfaction & $3.58 \pm 0.88$ & $4.42 \pm 0.78$ & 12.0 & $<< 0.01$** \\
\hline
\end{tabular}
\end{table}

We evaluated user experience across both conditions using ratings from mid-study and post-study assessments. As shown in Table \ref{tab:user-experience}, CareerPooler significantly outperforms ChatGPT on engagement ($p<<0.01$) and satisfaction ($p<<0.01$). The qualitative analysis reveals that CareerPooler's gamification mechanisms create multiple reinforcing factors including progressive information disclosure, physical control and agency, immersive time experience, curiosity-driven exploration, multimedia enhancement, and reduced psychological pressure that collectively enhance user experience and learning outcomes.

\begin{table}[h]
\centering
\caption{RQ1 - Engagement: CareerPooler significantly enhances user engagement.}
\label{tab:rq1-engagement}
\begin{tabular}{p{0.23\textwidth} p{0.75\textwidth}}
\hline
\multicolumn{2}{l}{\textbf{RQ1 - Engagement}} \\
\hline
\textbf{Mechanism} & \textbf{Participant Quotes} \\
\hline
Progressive Information Disclosure Enhances Comprehension &
P4: ``Each event's generation took effort, so I read more carefully.'' \newline
P21: ``It is step by step, so it's useful.'' / ``Seeing things step by step allows you to build a storyline.'' \newline
P24: ``ChatGPT's content is like a research paper, while CareerPooler delivers it step by step, so I read carefully.'' / ``CareerPooler's information prompts stimulate my curiosity and open up my thinking.'' \\
\hline
Physical Control and Agency &
P4: ``The process of hitting balls gives a feeling of controlling my own future.'' \newline
P13: ``CareerPooler lets you choose what to do next and how many days to spend.'' \newline
P20: ``CareerPooler allows you to choose what you want to do based on information prompts; ChatGPT doesn't give choices.'' \\
\hline
Immersive Time Experience and Urgency &
P11: ``Knowing two years of results in one evening feels precious---I learned time management.'' \newline
P12: ``You will have that feeling, and it will have a sense of time urgency.'' \newline
P20: ``CareerPooler really made me experience two years.'' / `ChatGPT's timeline feels indefinitely far away, like nothing has even started.'' \\
\hline
Multimedia Enhancement and Emotional Connection &
P10: ``Images can provide some emotion.'' \newline
P13: ``I really like the real-time generated images, they feel very realistic.'' \\
\hline
Intrinsic Motivation and Sustained Appeal &
P4: ``Even after the experiment ends, when I'm confused about the future, I would play this game.'' \newline
P11: ``A very moving pool game.'' \newline
P16: ``It has both randomness and gives you a feeling of a good life, making you unable to stop playing.'' \newline
P24: ``CareerPooler is more experiential than ChatGPT, with stronger empathy, providing some emotional support and practical future advice.'' \\
\hline
Effort Paradox: \newline High Operational Demand, Low Perceived Burden &
P3: ``Although I performed more operations in CareerPooler, it felt like playing a game, unlike ChatGPT where I constantly had to think about what questions to ask.'' / ``ChatGPT feels very stressful, like the neighbor's perfect child, making you feel it's unachievable.'' \newline
P20: ``CareerPooler has lighter burden; ChatGPT is too serious.'' \newline
P24: ``Career planning is quite a serious matter, but CareerPooler made the process more bearable.'' \\
\hline
\end{tabular}
\end{table}

\subsubsection{\textbf{RQ1 - Engagement: }CareerPooler's gamification mechanisms significantly enhance user engagement through multiple reinforcing factors (Table \ref{tab:rq1-engagement})}

The statistically significant difference in engagement ratings ($p << 0.01$) demonstrates how CareerPooler's gamification mechanisms effectively sustain user engagement throughout the career exploration process. Behavioral data corroborates these subjective ratings: 24 participants conducted 63 career exploration sessions with CareerPooler (avg 2.63 rounds per person), with 9/24 (37.5\%) actively requesting additional rounds, while ChatGPT was used only 43 times (avg 1.79 rounds per person) with no participants requesting continued use. Figure \ref{fig:ballanalysis} presents the performance progression across ball types and gaming sessions. Multiple participants described achieving immersive states that transcended the experimental context. P11 explained, ``Playing CareerPooler distracted me---I wouldn't think about playing pool but would immerse myself in my career experiences.'' Similarly, P12 highlighted the immersiveness of the gameplay, noting, ``The pool game is very immersive---every round, every game, choosing which ball to hit, how many days to use, every step requires thinking.'' This sense of absorption was further reinforced by P15, who stated, ``I was too immersed in playing pool, I forgot I was doing career planning.'' Table \ref{tab:rq1-engagement} shows how the combination of progressive information disclosure, physical agency, urgency induced by the timeline, and immersive gameplay mechanics fostered deeper engagement and sustained motivation compared to the linear chatbot baseline.

\subsubsection{\textbf{RQ1 - Information Gain:} CareerPooler provides comparable levels of information gain to ChatGPT, with trade-offs between accessibility and completeness (Table \ref{tab:rq1-infogain})}

Quantitative analysis revealed no statistically significant difference in information gain between CareerPooler and ChatGPT (W$=$29.0, $p=0.07$). While qualitative feedback revealed trade-offs between accessibility and completeness. Participants frequently characterized ChatGPT’s responses as overly complex and cognitively demanding, with 13/24 participants (54.1\%) explicitly citing information overload. In contrast, CareerPooler delivered information in smaller, digestible segments anchored to concrete events. This step-wise progression allowed participants to concentrate on individual decisions without becoming overwhelmed, even though it sometimes came at the expense of overall informational richness.

\begin{table}[h]
\centering
\caption{RQ1 - Information Gain: CareerPooler provides comparable levels of information gain to ChatGPT, reflecting trade-offs between accessibility and completeness.}
\label{tab:rq1-infogain}
\begin{tabular}{p{0.23\textwidth} p{0.75\textwidth}}
\hline
\multicolumn{2}{l}{\textbf{RQ1 - Information Gain}} \\
\hline
\textbf{Mechanism} & \textbf{Participant Quotes} \\
\hline
Trade-offs between Accessibility and Completeness: Step-by-step Processing with Fragmentation &
\textit{Advantage of CareerPooler} \newline
P6: ``CareerPooler shows information in little events that are broken up with fun between them, so it's not overwhelming and greatly reduces the mental overwhelm.'' \newline
P20: ``ChatGPT gives too much complexity, it's exhausting to read.'' \newline
P24: ``ChatGPT's content is like a research paper, while CareerPooler delivers it step by step, so I read carefully.'' \newline
\textit{Disadvantage of CareerPooler} \newline
P4: ``CareerPooler's key steps are somewhat limited compared to ChatGPT's information richness.'' \newline
P5: ``The information provided is too little.'' \newline
P8: ``Even though I hit 11 balls, the information seems quite limited because each one only gives you one point.'' \\
\hline
Simulated Career Progression Enhances Learning &
P11: ``Experiencing events happening to myself at low cost was quite shocking.'' \newline
P12: ``CareerPooler gave a feeling of truly learning something, a kind of lived-through understanding.'' \newline
P22: ``CareerPooler slowed down your thinking speed, allowing you to focus on thinking about each event.'' \\
\hline
Spatial Organization Reveals Career Relationships &
P9: ``In CareerPooler, I can see all possibilities laid out before me, making me want to try different combinations.'' \newline
P19: ``ChatGPT gives a single route, while CareerPooler gives me a feeling that life is multifaceted.'' \newline
P18: ``It's not something I can immediately use---I wouldn't be able to follow the plan step by step like ChatGPT suggests.'' (broad exploration sometimes came at the cost of implementation depth) \\
\hline
\end{tabular}
\end{table}

\begin{table}[h]
\centering
\caption{RQ1 - Authenticity: CareerPooler achieves comparable perceived authenticity through balanced narrative design.}
\label{tab:rq1-authenticity}
\begin{tabular}{p{0.23\textwidth} p{0.75\textwidth}}
\hline
\multicolumn{2}{l}{\textbf{RQ1 - Authenticity}} \\
\hline
\textbf{Mechanism} & \textbf{Participant Quotes} \\
\hline
Balanced Event Distribution Enhances Perceived Realism &
P6: ``The funding cut event felt too real---this actually happened to my lab colleague last year. The baseline system just kept giving me promotions and successes.'' \newline
P21: ``CareerPooler raises many real problems, telling me I'll encounter these issues... ChatGPT only gives very positive responses.'' \newline
P24: ``ChatGPT is too optimistic about reality, while CareerPooler provided many positive and negative events, feeling more aligned with reality.'' \\
\hline
Narrative Coherence Through Storytelling Mechanisms despite Illogical Sequences &
P10: ``CareerPooler is more immersive, able to show what I will experience.'' \newline
P18: ``CareerPooler is more like generating stories, more realistic in form.'' / ``The system sometimes has illogical sequences---I was promoted to senior engineer before completing my degree.'' \newline
P21: ``CareerPooler allows you to build a storyline step by step.'' \\
\hline
Randomness as Authentic Career Development Metaphor &
P17: ``CareerPooler is very life-like; life is very random.'' \newline
P18: ``Even when you want to fine-tune, there's still a kind of randomness.'' / ``Often you don't know where the white ball will drift.'' \newline
P19: ``This game's philosophy contains an element of randomness in effort... life is about not trying too hard, or you won't get through.'' \\
\hline
Physical Skill Requirements Create Authenticity-Accessibility Tension &
\textit{Advantage of CareerPooler} \newline
P12: ``Pool ball striking simulates the vitality of breaking through difficulties and growing upward, giving people confidence.'' \newline
\textit{Disadvantage of CareerPooler} \newline
P1: ``The difficulty of operating the mouse itself exceeded the difficulty of career development.'' \newline
P10: ``The pool balls are so hard to hit, how much I can experience this game depends on my pool playing skills.'' \newline
P16: ``At first I kept missing balls, worried that I would accomplish nothing.'' \\
\hline
Personalization Limitations Across Both Systems &
\textit{CareerPooler} \newline
P15: ``I gave relatively little information initially, so the generated content was also quite general, not very targeted to me personally.'' \newline
P18: ``Possibly because I didn't give him too many details, so what he gave me wasn't that close to what I was thinking, it was relatively generic.'' \newline
\textit{ChatGPT} \newline
P10: ``CareerPooler's information content is tailored, while ChatGPT's content doesn't feel targeted to me---I think this content could be given to anyone.'' \newline
P16: ``Some suggestions didn't actually consider where I live now, my past learning experiences, my past life experiences.'' \newline
P19: ChatGPT's approach was ``more like non-personalized career planning.'' \\
\hline
\end{tabular}
\end{table}

\subsubsection{\textbf{RQ1 - Authenticity: } CareerPooler achieves comparable perceived authenticity through balanced narrative design incorporating realistic setbacks (Table \ref{tab:rq1-authenticity})}

\label{result-authenticity}

While perceived authenticity ratings showed no statistically significant difference between CareerPooler and ChatGPT (W$=$39.5, $p=0.22$), qualitative analysis reveals distinct mechanisms through which each system establishes credibility and authenticity in career guidance.

At the same time, 14/24 (58.3\%) participants specifically emphasized the realism of the events, with 10 participants noting that negative outcomes in particular felt ``strikingly" authentic. As P7 explained, “This game is so realistic—much more authentic than ChatGPT. ChatGPT tends to be idealistic, offering universal pathways without much consideration for real-world factors. The pool game is more realistic because it considers many practical factors—you need to spend a lot of time building skills in the early stages.” Others drew direct parallels to their own experiences: P11 remarked that “playing pool slowly and not being skilled at first is like how people take detours when starting out in life,” while P12 noted that “efforts don’t always yield rewards—advisors changing directions, lab errors—these difficulties are authentic.” Table \ref{tab:rq1-authenticity} summarizes the factors shaping participants’ perceptions of CareerPooler’s authenticity.

On the other hand, 6/24 participants (25\%) raised concerns about coherence and temporal consistency, noting that "the events themselves seem authentic, but the timeline is somewhat disorganized." This limitation reflects the broader challenge of large language models in maintaining chronological reasoning, an issue we further discuss in Section~\ref{limitation:timeline-inconsistency}. Notably, participants did not limit themselves to exploring realistic career paths. Owing to CareerPooler's playful and game-like qualities, a substantial minority (24/63, 38\%) also engaged in imaginative trajectories, experimenting with aspirational or unconventional roles such as becoming a forest ranger, Oscar-winning actor, or even a Pokémon trainer. These imaginative explorations broaden the notion of authenticity, framing it not only as the replication of real-world experiences but also as the construction of plausible and logically coherent career possibilities.

\begin{table}[h]
\centering
\caption{RQ1 - Satisfaction: CareerPooler significantly outperforms ChatGPT in user satisfaction.}
\label{tab:rq4-qual}
\begin{tabular}{p{0.23\textwidth} p{0.75\textwidth}}
\hline
\multicolumn{2}{l}{\textbf{RQ1 - Satisfaction}} \\
\hline
\textbf{Mechanism} & \textbf{Participant Quotes} \\
\hline
Emotional Connection and Companionship &
\textit{CareerPooler} \newline
P11: ``A very moving game.'' \newline
P12: Experienced ``a sense of companionship and enticing you to learn this skill.'' \newline
P24: ``CareerPooler is more experiential than ChatGPT, with stronger empathy, providing some emotional support and practical future advice.'' / ``When CareerPooler's stories closely resembled my personal experiences, I felt very immersed.'' \newline
\textit{ChatGPT} \newline
P10: ``ChatGPT's text feels somewhat cold.'' \newline
P12: ``(ChatGPT is) very objective, like a God's-eye-view teacher telling you, very rational.'' \\
\hline
Achievement and Control Through Interactive Decision-Making &
P1: ``As a game, hitting pool balls is very satisfying.'' \newline
P4: ``The process of hitting balls gives a feeling of controlling my own future.'' \newline
P5: ``I really like hitting milestone balls because there's a milestone count record on the right side---I want to fill up that number.'' \newline
P13: ``CareerPooler lets you choose what to do next and how many days to spend.'' \newline
P20: ``My pool technique is also improving, which gives me a sense of achievement.'' \\
\hline
Game Metaphor Reduces Psychological Pressure and Enables Creative Exploration &
\textit{CareerPooler} \newline
P3: ``It feels like a game where I can try crazy things, like completely switching fields, without real consequences.'' / ``Although I performed more operations in CareerPooler, it felt like playing a game.'' \newline
P12: ``Life paths become increasingly narrow, so we really need software like this.'' \newline
\textit{ChatGPT} \newline
P3: ``ChatGPT feels very stressful, like the neighbor's perfect child, making you feel it's unachievable.'' \newline
P7: ``ChatGPT plans very well, but whether I can achieve it is another matter.'' \newline
P16: ``It gives detailed career planning that makes me feel I can take it step by step, but also makes me feel tremendous pressure and trouble.'' \\
\hline
\end{tabular}
\end{table}

\subsubsection{\textbf{RQ1 - Satisfaction: } CareerPooler significantly outperforms ChatGPT in user satisfaction (Table \ref{tab:rq4-qual})}

CareerPooler achieved statistically significant higher satisfaction ratings than ChatGPT (M$=$4.42 vs 3.58, $p<<0.01$), emerging from multiple interconnected factors that transformed career exploration from a stressful analytical task into an engaging exploratory experience. This satisfaction advantage reflected differences in how the systems balanced psychological comfort with informational utility.
\subsection{RQ2: CareerPooler improves clarity in career exploration outcomes}

\begin{table}[h]
\centering
\caption{Career exploration outcome changes under ChatGPT and CareerPooler conditions. Significance tests compare pre-study baselines with post-condition scores under each system using Wilcoxon signed-rank tests. Results indicate that CareerPooler produced a significant improvement in Career Clarity ($p < 0.05$), whereas neither system showed significant changes in Step-wise Understanding or Career Confidence. The order of system exposure was counterbalanced across participants.}
\label{tab:career-comparison}
\resizebox{\textwidth}{!}{%
\begin{tabular}{lcccccccc}
\hline
\multirow{2}{*}{\textbf{Dimension}} & \multirow{2}{*}{\textbf{Pre-study}} & \multicolumn{3}{c}{\textbf{ChatGPT}} & \multicolumn{3}{c}{\textbf{CareerPooler}} \\
\cline{3-5} \cline{6-8}
 & & \textbf{Post} & \textbf{W} & \textbf{p-value} & \textbf{Post} & \textbf{W} & \textbf{p-value} \\
 & $(M \pm SD)$ & $(M \pm SD)$ & & & $(M \pm SD)$ & & \\
\hline
Career Clarity & $3.21 \pm 0.83$ & $3.54 \pm 0.88$ & 41.5 & 0.28 & $3.92 \pm 1.02$ & 17.0 & 0.01* \\
Step-wise Understanding & $3.38 \pm 0.97$ & $3.50 \pm 0.83$ & 63.0 & 0.79 & $3.88 \pm 1.04$ & 45.0 & 0.06 \\
Career Confidence & $3.38 \pm 1.01$ & $3.75 \pm 0.94$ & 43.5 & 0.19 & $3.79 \pm 0.88$ & 30.0 & 0.07 \\
\hline
\end{tabular}%
}
\begin{tablenotes}
\small
\item Note: * $p < 0.05$. W and p-values represent Wilcoxon signed-rank test results comparing pre-study with post-condition scores.
\end{tablenotes}
\end{table}

We compared pre-study baseline measurements with post-condition assessments to evaluate changes in career exploration outcomes. Results showed that only CareerPooler produced significant improvement in career clarity (W$=$17.0, $p=0.01$), increasing from 3.21 pre-study to 3.92 post-study. In contrast, career clarity improvement under ChatGPT conditions was not significant (from 3.21 to 3.54, $p=0.28$). The qualitative analysis reveals that CareerPooler enhances career exploration outcomes through experience-based learning and reflective processing, step-wise understanding via progressive disclosure, and confidence building through balanced preparation and agency enhancement.

\begin{table}[h]
\centering
\caption{RQ2 - Career Clarity: CareerPooler significantly improves career clarity through experience-based learning and reflective processing.}
\label{tab:rq2-clarity}
\begin{tabular}{p{0.23\textwidth} p{0.75\textwidth}}
\hline
\multicolumn{2}{l}{\textbf{RQ2 - Career Clarity}} \\
\hline
\textbf{Mechanism} & \textbf{Participant Quotes} \\
\hline
Reflective Processing Through Experience-Based Interaction &
P11: ``Low-cost experience of events happening to oneself was quite shocking.'' \newline
P12: ``CareerPooler triggers my reflection and impact, readjusting myself.'' / ``A feeling of really learning something, a sense of 'lived-through understanding'.'' / ``CareerPooler focuses more on the growth process, accompanying you step by step to progress.'' \newline
P22: ``CareerPooler slowed down your thinking speed, allowing you to focus on thinking about each event.'' \\
\hline
Physical Interaction Creates Stronger Mental Models &
P11: ``Playing pool slowly and not well is like taking some detours at the beginning of life.'' \newline
P13: ``During the pool process, it triggers some thinking.'' \newline
P19: ``It enhances the experiential nature of career planning, making me more actively explore.'' \\
\hline
Sequential Ball Interactions Reveal Causal Career Relationships &
P4: ``CareerPooler generated experiences, demonstrating what skills are useful for and the causal relationships between events.'' \newline
P14: ``Seeing how one skill ball unlocks multiple milestone opportunities made career step connections clearer.'' \newline
P19: ``CareerPooler may not help much with career planning, but it gives experienced things and leaves the feeling of experiencing a career.'' \newline
P21: ``CareerPooler allows you to see things step by step and build a storyline.'' \\
\hline
\end{tabular}
\end{table}

\subsubsection{\textbf{RQ2 - Career Clarity:} CareerPooler significantly improves career clarity through experience-based learning and reflective processing (Table \ref{tab:rq2-clarity})} Participants reported that CareerPooler improved their career clarity by fostering reflective processing, strengthening mental models through embodied interaction, and revealing causal relationships between career events. The experience-based design encouraged users to slow down and reflect on each event, leading to a ``lived-through understanding'' of career development (e.g., P11, P12, P22, P23). Physical interactions with the pool interface further anchored abstract insights in tangible experiences, enhancing self-awareness and motivating active exploration (P11, P13, P19). Sequential ball interactions helped participants perceive causal links between skills, milestones, and outcomes, enabling them to construct coherent storylines of career progression (P4, P14, P21). While some noted that CareerPooler is more exploratory than prescriptive (P19), the integration of reflective, embodied, and sequential mechanisms collectively facilitated clearer mental models of career pathways.

\subsubsection{\textbf{RQ2 - Step-wise Understanding:} Supporting step-wise career understanding through progressive disclosure and causal relationship visualization}

While participants showed improvements in understanding steps needed to achieve goals under both conditions, these improvements did not reach statistical significance. Nevertheless, qualitative feedback indicated that the step-wise progression appeared to help participants develop clearer understanding of goal achievement processes. Participants described enhanced clarity: ``I know I have a clear goal, and now through playing this game, do I know what specific steps I need to take before achieving this goal. And do I know these steps very clearly.'' (P1)

\begin{table}[h]
\centering
\caption{RQ2 - Career Confidence: CareerPooler builds career confidence through balanced preparation and agency enhancement.}
\label{tab:rq2-confidence}
\begin{tabular}{p{0.23\textwidth} p{0.75\textwidth}}
\hline
\multicolumn{2}{l}{\textbf{RQ2 - Career Confidence}} \\
\hline
\textbf{Mechanism} & \textbf{Participant Quotes} \\
\hline
Balanced Expectations Create Realistic Confidence &
P6: ``Experienced some skill and random events in the pool interface and found they have relevant experience, thus became confident.'' \newline
P16: ``After playing, I feel like I've become stronger.'' \newline
P24: ``Clarified what I can do and what depends on luck, can be more confident about my career planning.'' \\
\hline
Resilience Preparation Through Setback Simulation &
P6: ``The failures CareerPooler generated are important for preparing younger students' resilience mindset.'' \newline
P7: ``Seeing that even successful paths include failures made me feel better about my own failures. ChatGPT just kept telling me to `stay positive' without acknowledging that failure is normal.'' \newline
P22: ``Random event balls made a very deep impression, with success and failure events, making me think that I should also be prepared to face possible failures.'' \\
\hline
Agency and Control Enhance Self-Efficacy &
\textit{CareerPooler} \newline
P4: ``The process of hitting balls gives a feeling of controlling my own future.'' \newline
P20: ``CareerPooler allows you to choose what you want to do based on information prompts (information hints); ChatGPT doesn't give choices.'' \newline
\textit{ChatGPT} \newline
P6: ``ChatGPT listed many skills not yet mastered, making people feel not so confident.'' \\
\hline
\end{tabular}
\end{table}

\subsubsection{\textbf{RQ2 - Career Confidence: }CareerPooler builds career confidence through balanced preparation and agency enhancement (Table \ref{tab:rq2-confidence})}

CareerPooler appeared to enhance participants' career confidence through multiple mechanisms that balanced realistic preparation with empowerment, contrasting with ChatGPT's approach that provided comprehensive guidance but potentially undermined self-efficacy.
\subsection{RQ3: Participants demonstrate varying levels of awareness regarding AI's impact on career development}

Interview analysis reveals that participants demonstrate diverse awareness levels and adaptation strategies regarding AI's career impact. Technical professionals show more nuanced understanding of AI's implications, while students exhibit varying degrees of awareness despite limited industry experience. Four distinct adaptation strategies emerged: proactive integration, tool utilization, defensive avoidance, and traditional path dependency, reflecting different approaches to navigating AI-influenced career environments.

\subsubsection{\textbf{RQ3 - AI Awareness: } Diverse awareness levels reveal AI's uneven career impact across professional contexts (Table \ref{tab:rq3-awareness})}

Interview analysis revealed varying levels of awareness regarding AI's impact on careers. Among 24 participants, most participants reported knowing someone whose career was affected by AI automation. However, their understanding of AI capabilities differed based on their professional proximity to AI technology.

\begin{table}[h]
\centering
\caption{RQ3 - AI Awareness: Diverse awareness levels reveal AI's uneven career impact across professional contexts.}
\label{tab:rq3-awareness}
\begin{tabular}{p{0.23\textwidth} p{0.75\textwidth}}
\hline
\multicolumn{2}{l}{\textbf{RQ3 - AI Awareness}} \\
\hline
\textbf{Mechanism} & \textbf{Participant Quotes} \\
\hline
Technical Professionals Demonstrate Nuanced Understanding &
P12: ``I previously started a business doing IP-customized clothing...since AI came out, partners can save a lot of money.'' \newline
P15: ``This year looking for jobs, SWE position demand has decreased, and salaries offered aren't high.'' \newline
P18: ``The most anxious are the SDE people around me, but I don't think the impact on them will be the greatest.'' \\
\hline
Students Show Varying Awareness Despite Limited Industry Experience &
P17: ``HR uses AI to screen resumes, reducing a lot of time spent screening resumes.'' \newline
P20: ``Many PhD students graduate thanks to AI.'' \newline
P21: ``People around me use AI to easily make prototypes.'' \\
\hline
AI Impact Creates Career Uncertainty and Strategic Adaptation &
P10: ``Although AI has lowered barriers, it's brought me more worry and uncertainty.'' / ``Although the threshold for entering the industry has become smaller, there's still a big gap with industry professionals.'' \newline
P13: ``Worried about whether I'll be unemployed in the future.'' \newline
P18: ``I think it has greatly affected the next generation, education has lag.'' \\
\hline
\end{tabular}
\end{table}

\begin{table}[h]
\centering
\caption{RQ3 - AI Strategies: Participants employ diverse strategies for integrating AI considerations into career planning.}
\label{tab:rq3-strategies}
\begin{tabular}{p{0.23\textwidth} p{0.75\textwidth}}
\hline
\multicolumn{2}{l}{\textbf{RQ3 - AI Strategies}} \\
\hline
\textbf{Mechanism} & \textbf{Participant Quotes} \\
\hline
Proactive Integration Strategy &
P12: ``Avoid competing with AI, quickly move up, become someone who integrates resources and discovers problems.'' \newline
P18: ``AI's emergence made me choose to research AI---if you can't beat them, join them.'' \\
\hline
Tool Utilization Strategy &
P15: ``I think AI is a productivity tool, an indispensable tool.'' \newline
P18: ``AI has leveled many knowledge gaps.'' / ``Now everyone publishes papers and writes code quickly, AI's positive impact.'' \newline
P20: ``I work in AI so AI can help me with some basic tasks like grammar checking and basic code writing.'' / ``Overall I think AI is a very useful tool.'' \\
\hline
Defensive Avoidance Strategy &
P8: ``Music AI suno is very powerful, but musicians are still optimistic, augmentation.'' \newline
P17: ``The path is walked step by step by oneself, AI is not really in my career considerations.'' \newline
P21: ``AI will replace many basic skills.'' / ``Human needs for food, drink, and basic necessities will always be here, so we shouldn't focus too much on polishing skills, but should pay more attention to human desires themselves.'' \\
\hline
\end{tabular}
\end{table}

\subsubsection{\textbf{RQ3 - AI Strategies: }Participants employ diverse strategies for integrating AI considerations into career planning (Table \ref{tab:rq3-strategies})}

Content analysis of participants' career exploration revealed three understanding and coping strategies regarding AI's impact on career development: proactive integration, tool utilization, and defensive avoidance. Corresponding quotes are provided in Table \ref{tab:rq3-strategies}.

%% file: sections/6-discussion_conclusion.tex
\section{Discussion}

\subsection{Summary of Results}

Our comparative study reveals that CareerPooler significantly outperforms conversational AI (e.g. ChatGPT) across multiple dimensions of user experience and outcomes. For \textbf{RQ1 - User Experience}, CareerPooler demonstrates statistically significant advantages in engagement ($p<<0.01$) and overall satisfaction ($p<<0.01$) through gamification mechanisms including progressive information disclosure, physical control and agency, immersive time experience, and multimedia enhancement. The experiential approach reduces information overload while creating deeper engagement through spatial reasoning tasks that make abstract career relationships tangible and interactive.

For \textbf{RQ2 - Outcome}, CareerPooler improves career clarity ($p=0.01$), while showing no significant improvement in understanding of step-by-step career progression and confidence in implementation. Qualitative findings suggest that the system’s step-wise progression and embodied interaction foster stronger mental models, enabling participants to better grasp the causal relationships between career events and to form a clearer understanding of how goals can be achieved over time. Moreover, the balanced integration of both positive and negative career scenarios provides more realistic preparation and cultivates resilient confidence by simulating potential setbacks.

For \textbf{RQ3 - AI Awareness}, participants demonstrate diverse awareness levels regarding AI's career impact and employ three distinct adaptation strategies: proactive integration, tool utilization, and defensive avoidance. Technical professionals show more nuanced understanding of AI implications, while students exhibit varying awareness levels given their limited industry experience. These findings reveal that AI's career impact creates both opportunities and anxieties, highlighting the need for adaptive career exploration tools that accommodate diverse user attitudes toward AI integration.


\subsection{Modeling Abstract Real-world Events Through Spatial Game Mechanics}

The design of CareerPooler illustrates a generalizable interaction paradigm: leveraging simple spatial operations and metaphorical mappings to transform abstract, hard-to-grasp real-world events into forms that can be understood, manipulated, and experienced. In this system, actions such as striking balls, calculating angles, and predicting trajectories—originally the mechanics of a pool game—are reinterpreted as metaphors for individual choices, attempts, and risk management in career development. This metaphorical interaction not only concretizes the abstract notion of ``career pathways'', but also makes causal relationships and uncertainties more tangible through the spatial distribution and non-linear trajectories of event balls. Our empirical findings provide evidence for how this mechanism operates across multiple dimensions of user experience and career exploration outcomes.

\subsubsection{From Engagement to Meaning-making Through Embodied Interaction}

The engagement and satisfaction advantages observed in our study ($p<<0.01$ for both) suggest that spatial game mechanics do more than entertain—they structure how users process and internalize abstract information. Participants described absorption states where career exploration became intrinsically motivating rather than cognitively burdensome, consistent with flow theory~\cite{csikszentmihalyi1990flow}. This engagement was not superficial: the ``slowing down'' effect reported by participants (e.g., P22: ``CareerPooler slowed down your thinking speed'') corresponds to the reflective play mechanisms identified by Miller et al.~\cite{miller2024reflective}, where deliberate pacing creates cognitive space for processing disruptive experiences. The satisfaction participants derived from interactive decision-making—choosing which ball to strike, calibrating force, observing outcomes—reflects what Salen and Zimmerman~\cite{salen2003rules} term meaningful play: a legible relationship between player action and system outcome that sustains both motivation and comprehension.

These findings indicate that the pool metaphor functions as what Barr et al.~\cite{barr2002taxonomy} call a process metaphor: it does not merely label career elements but explains their dynamic relationships through spatial interaction. As participants struck balls and observed resulting trajectories and collisions, they were constructing mental models through embodied reasoning about cause, consequence, and contingency. This experiential mode of sense-making supports Bogost's~\cite{bogost2007persuasive} argument that procedural systems convey complex arguments more effectively than declarative statements—precisely because they require the user to enact rather than merely read the underlying logic. When a participant's carefully aimed shot is deflected by an unexpected collision, the system argues that deliberate effort can be redirected by forces outside one's control. When a ball narrowly misses a pocket, it argues that proximity to a goal does not guarantee achievement. These are conclusions players draw from their own experience with the system's rules, not messages delivered through text.

\subsubsection{Multiple Play Modes and the Role of Balanced Valence}

In Hunicke et al.'s~\cite{hunicke2004mda} MDA framework, CareerPooler's mechanics (pool physics, time-as-distance, ball categories) give rise to dynamics (emergent career narratives, strategic-vs-exploratory play styles, unexpected event chains) that produce aesthetics of discovery, challenge, and narrative. The fact that different participants gravitated toward different play styles—some methodically targeting milestone balls, others embracing random collisions, and 38\% exploring fantastical careers such as becoming a forest ranger or Pok\'{e}mon trainer~\cite{caillois1961man}—suggests that the system simultaneously supports multiple modes of meaningful play~\cite{salen2003rules}.

The resilience-building effect of setback simulation further extends the serious games literature~\cite{wouters2013meta}. Participants reported that balanced positive and negative events enhanced perceived authenticity and built confidence (P6, P7, P22), demonstrating that balanced valence in AI-generated content—rather than uniformly positive narratives—can itself be a design resource within playful, low-stakes contexts~\cite{to2022interactive, fernandez2021climate}. This finding has particular relevance for systems combining LLMs with game-based interaction, given that large language models tend to overrepresent successful outcomes due to survivorship bias in training data.

\subsubsection{Generalizability: A Metaphor-to-Mechanics Design Method}

This metaphorical mechanism offers three avenues for generalization. First, it provides a way to model other complex social processes. Domains such as financial decision-making and educational trajectories share characteristics of uncertainty, path dependency, and staged goals. Spatial metaphorical interactions may help users better grasp the dynamic properties of these processes. Second, it emphasizes the socio-psychological significance of the ``learning curve'': early failures and repeated attempts are not merely technical obstacles but serve as metaphors for real-world growth and resilience. As P11 reflected, ``Shooting balls is difficult at first, like life itself, but becomes smoother with mastery'', capturing the shift from early setbacks to eventual competence.

Third, and most relevant to the game design community, CareerPooler demonstrates a metaphor-to-mechanics translation process that other designers can adapt: (1) identify the core properties of the target domain (for career exploration: effort, uncertainty, path dependency, partial information); (2) select a game whose inherent mechanics naturally embody those properties (pool: force, collision, trajectory, hidden outcomes); (3) formalize the mapping between game elements and domain concepts; and (4) use generative AI to populate the mechanics with domain-specific content. This process is not specific to career exploration—it offers a reusable design method for any serious game aiming to model abstract real-world processes through spatial interaction.

Nevertheless, the generalization of metaphorical interactions faces challenges. Users vary in their spatial reasoning abilities, and if a system relies too heavily on operational skills, some individuals may focus more on the ``gameplay'' rather than its metaphorical meaning. This risks diminishing the system's value as a model of abstract real-world events. Future iterations should therefore incorporate adaptive difficulty mechanisms, enabling users to adjust the complexity of shooting mechanics according to their own abilities while preserving the core metaphorical relationship to real-world challenges.

\subsection{Design Implications for AI-powered Career Exploration System}

Considering that there are currently few GenAI-powered career exploration tools \cite{jeonLettersFutureSelf2025,du2024careersim}, and that those which exist are primarily based on conversational interfaces with limited interactivity, our work offers five design implications for the development of AI-powered career exploration systems and, more broadly, for interactive educational tools:

\begin{itemize}
\item \textbf{Leveraging experience-based learning alongside conversational interfaces:} The enhanced engagement and information retention achieved through hands-on interaction versus traditional conversational approaches suggests that experience-based learning paradigms should be prioritized in educational AI systems. Our results provide empirical evidence for what the serious games literature has long argued theoretically~\cite{wouters2013meta, taly2017serious}: active engagement through structured interaction produces deeper learning than passive information consumption. This finding has implications for diverse applications such as skill development platforms.

\item \textbf{Use spatial metaphors to ground abstract concepts:} The effectiveness of the pool table metaphor in making connections between skills, milestones, and outcomes tangible demonstrates how spatial representations can act as powerful cognitive scaffolds for abstract decision-making processes. Grounded in Lakoff and Johnson's~\cite{lakoff1980metaphors} conceptual metaphor theory, our pool-table mapping demonstrates that spatial metaphors do more than visualize information—they enable users to reason about abstract relationships through embodied spatial cognition. This approach can extend to other complex domains such as financial planning, educational pathways, or organizational development.

\item \textbf{Employ progressive disclosure to manage complexity:} CareerPooler's progressive disclosure approach demonstrates that breaking down complex information into manageable, contextually relevant segments reduces cognitive overload and supports deeper engagement. This mechanism leverages what Costikyan~\cite{costikyan2013uncertainty} identifies as hidden information uncertainty to sustain curiosity-driven exploration~\cite{to2016integrating}. This principle should be applied to other domain-specific AI systems where information overload poses challenges.

\item \textbf{Time rationality in AI-generated narratives:} The timeline inconsistencies observed in LLM-generated career scenarios underscore the critical importance of maintaining logical temporal sequences in AI systems that generate life narratives or sequential events.

\item \textbf{Balance valence in AI-generated game content:} Our finding that balanced positive and negative events enhanced perceived authenticity and built resilience (Section~\ref{result-authenticity}) has direct implications for any system combining LLMs with game-based interaction. Deliberate inclusion of setbacks, failures, and neutral outcomes is not merely a realism feature but an active design resource that supports emotional preparation and realistic expectation-setting~\cite{fernandez2021climate}.
\end{itemize}

\subsection{Ethical Considerations}

\label{sec:ethical_considerations}

Career exploration inherently involves ethical sensitivity, as it relates to identity formation, self-efficacy, and future aspirations. Simulated trajectories may influence participants’ perceptions of their own potential and risks. To address this, we carefully designed the CareerPooler system aiming to provide useful and constructive career information to participants, while balancing positive and negative events to reflect realistic experiences. We also anonymized all data and emphasized participants’ autonomy, including the option to withdraw at any time. Participants provided positive feedback, with some even describing the system as “a very moving game,” suggesting that our design fostered both emotional resonance and reflective learning. By foregrounding well-being, privacy, and supportive exploration, we sought to ensure that the system benefits participants without imposing undue pressure or reinforcing anxiety.
\subsection{Limitation and Future Work}

Although the study indicates that CareerPooler holds promise for supporting career exploration, several limitations remain, offering potential directions for future research and system refinement.

\subsubsection{Confounding variables in system comparison}

Our comparative design evaluates CareerPooler as a holistic system against a conversational baseline, but CareerPooler simultaneously introduces multiple novel elements—gamification, spatial metaphor, progressive information disclosure, and AI-generated visual content. The current study cannot isolate which specific component drives the observed improvements in engagement, satisfaction, and career clarity. Future work should conduct ablation studies that systematically vary individual design elements (e.g., comparing a text-only pool interface with the full multimedia version, or a non-spatial gamified interface with the pool metaphor) to disentangle the contributions of each component.

\subsubsection{Accessibility and learning curve challenges}

Several participants struggled with ball-pocketing mechanics during initial sessions, indicating that the learning curve may create barriers to engagement. As P1 noted, ``The difficulty of operating the mouse itself exceeded the difficulty of career development,'' and P10 observed that ``how much I can experience this game depends on my pool playing skills.'' While some participants found metaphorical value in the initial difficulty—P11 remarked that ``playing pool slowly and not well is like taking some detours at the beginning of life''—the gameplay complexity sometimes overshadowed the career exploration objectives. Future iterations should implement adaptive difficulty systems, enabling users to adjust shooting mechanics to their skill level while preserving the core metaphorical learning experience.

\subsubsection{Inconsistency in AI-generated timeline}
\label{limitation:timeline-inconsistency}

Our study identified temporal inconsistencies in AI-generated career scenarios, where events occurred in illogical sequences. As P18 noted, ``The system sometimes has illogical sequences—I was promoted to senior engineer before completing my degree." These timeline errors undermine the authenticity and believability of the career exploration experience. Future work should investigate temporal reasoning improvements in large language models, particularly for generating life narratives that maintain logical chronological progression. This could involve developing specialized temporal validation mechanisms or incorporating explicit timeline constraints in AI-generated content.

\subsubsection{Privacy and trust concerns}
5/24 participants expressed privacy concerns about using AI for career exploration. P8 worried: ``I worry about privacy leaks, worried ChatGPT would leak my personal information." P17 explained their reservation: ``I think career planning is very personal... I'm unwilling to share so much of myself with AI, because AI is ultimately backed by a company." These concerns significantly affected usage behavior, with P18 noting: ``After all, it's a tool external to the company, can't leak all information to ChatGPT."

Interestingly, CareerPooler's gamified interface appeared to alleviate some privacy concerns, possibly because the interaction felt more like gameplay than data collection. However, this raised deeper concerns about AI dependency, with P8 reflecting: "Dependence on AI has made me lose intuition and judgment." Future work should address these trust concerns through transparent data handling practices, local processing options, and mechanisms that preserve user agency in the career exploration process.

\subsubsection{Long-term engagement and novelty effect}

While CareerPooler demonstrated higher short-term engagement compared to conversational AI, longitudinal studies are needed to assess sustained usage patterns and long-term career development outcomes. The engagement advantage observed in our study may partly reflect a novelty effect—participants were encountering a game-based career exploration interface for the first time, and the unfamiliarity itself may have contributed to heightened curiosity and absorption. Whether these engagement benefits persist after repeated use, or diminish as the novelty wears off, remains an open question. Future research should investigate how periodic engagement with career exploration tool influences actual career decisions and professional development trajectories over extended periods.



%% file: sections/7-appendix.tex
\bibliographystyle{ACM-Reference-Format}
\bibliography{reference}

\appendix

\section{Appendix A: Fine-tuning Data Example and Prompts}
\label{appendix:a}

\begin{tcolorbox}[breakable, colback=gray!5, colframe=gray!75!black, title=Fine-tuning Data Example, fonttitle=\bfseries]
\ttfamily\footnotesize
\begin{flushleft}
\{ \\
\ \ \ {"instruction"}: "As a career planner, you can generate a major career development event that the user will encounter in the next four months, along with three random events, three skills they may learn, and corresponding information hints, based on the user's self-introduction, past experiences, and current situation. Output the results in JSON format.",\\

\ \ \ {"input"}: "User's profile: I am a first-year master's student majoring in HCI in the United States, and I am very interested in human-centered AI. I hope to become a PhD student in human-computer interaction in the United States in the next two years. Current time:2024/10/28.\\

You are a career planner generating milestone<currentMilestoneNum>, three new random events, and three new skills based on:\\
User's profile All previously pocketed events Your output must be a strictly JSON formatted object containing: { "bigEvent<currentMilestoneNum>":"title: [Event Title] | [detailed description of the major career milestone that LOGICALLY FOLLOWS from all previous events]", "randomEvent<currentMilestoneNum>-1":"title: [Event Title] | [detailed event description that is DIRECTLY RELATED to bigEvent<currentMilestoneNum>, explaining what happens and its impact] [Positive/Neutral/Negative/Change: direction label]", "randomEvent<currentMilestoneNum>-1-hint":"[2-6 word extremely vague and ambiguous hint that does NOT reveal whether the event is positive, negative, or neutral]", "randomEvent<currentMilestoneNum>-2":"title: [Event Title] | [detailed event description that is DIRECTLY RELATED to bigEvent<currentMilestoneNum>, explaining what happens and its impact] [Positive/Neutral/Negative/Change: direction label]", "randomEvent<currentMilestoneNum>-2-hint":"[2-6 word extremely vague and ambiguous hint that does NOT reveal whether the event is positive, negative, or neutral]", "randomEvent<currentMilestoneNum>-3":"title: [Event Title] | [detailed event description that is DIRECTLY RELATED to bigEvent<currentMilestoneNum>, explaining what happens and its impact] [Positive/Neutral/Negative/Change: direction label]", "randomEvent<currentMilestoneNum>-3-hint":"[2-6 word extremely vague and ambiguous hint that does NOT reveal whether the event is positive, negative, or neutral]", "skill<currentMilestoneNum>-1":"title: [Skill Name] | [detailed description of this skill relevant to bigEvent<currentMilestoneNum>, what it involves, and how it helps current career stage]", "skill<currentMilestoneNum>-1-hint":"[2-6 word mysterious hint that doesn't clearly indicate what skill this is]", "skill<currentMilestoneNum>-2":"title: [Skill Name] | [detailed description of this skill relevant to bigEvent<currentMilestoneNum>, what it involves, and how it helps career progression]", "skill<currentMilestoneNum>-2-hint":"[2-6 word mysterious hint that doesn't clearly indicate what skill this is]", "skill<currentMilestoneNum>-3":"title: [Skill Name] | [detailed description of this skill relevant to bigEvent<currentMilestoneNum>, what it involves, and how it helps future development]", "skill<currentMilestoneNum>-3-hint":"[2-6 word mysterious hint that doesn't clearly indicate what skill this is]" }\\

Output requirements:\\
Format MUST STRICTLY be: "title: [title] | [detailed content description] [information label]" The format has THREE parts with specific structure: Title section: "title: [specific title]" Content section: " | [detailed content description]" Label section (for random events only): " [sentiment label]" Content descriptions MUST be detailed (at least 2-3 sentences) and cannot be omitted "bigEvent<currentMilestoneNum>" should be generated based on ALL pocketed events up to this point ALL random events and skills MUST be DIRECTLY RELATED to bigEvent<currentMilestoneNum> Skills should be specific abilities relevant to the current career stage with detailed descriptions Each randomEvent MUST end with label in square brackets: [Positive], [Neutral], [Negative], or [Change: direction] CRITICAL HINT REQUIREMENTS: -- ALL hints MUST be extremely vague, ambiguous, and mysterious -- NEVER use words that reveal the nature of the event or skill -- Use cryptic phrases like: "whispers emerge", "fog lifts slowly", "unknown beckons", "silence speaks", "doors appear", "winds shift" -- Hints should create curiosity without giving away whether something is beneficial or detrimental ALL bigEvents MUST occur within a TWO-YEAR timeframe from the start <acceptedChangesStr> NEVER omit any part of the required format Output only valid JSON User's profile: <userIntro> All pocketed events so far: <pocketedEvents> Past experiences: <pastExperiencesStr or 'None yet'> Accepted direction changes: <acceptedDirectionChanges> Current time: <currentTime>",\\

\ \ \ {"output"}: "{ "bigEvent1": "title: Enroll in HCI Master's | Your first step is going to the United States to study HCI for a master's degree, and soon you start taking courses such as Introduction to HCI, User Research, and UI Prototyping.", "randomEvent1-1": "title: Homesick | You find it hard to adapt to the United States, miss your family and friends, and become unmotivated. As a result, your grades in your first semester drop and you fail two courses. [Negative]", "randomEvent1-1-hint": "air thickens, energy wanes", "randomEvent1-2": "title: Graduate Satisfied | You do well in your courses, confirming your interest in HCI, and graduate with satisfaction. [Positive]", "randomEvent1-2-hint": "clear sky, first rays shine", "randomEvent1-3": "title: Become Interested in AR/VR | After taking a course in AR/VR design, you realize that this is what you want to do and hope to become an AR/VR designer in the future. [Change: HCI → AR/VR]", "randomEvent1-3-hint": "stars glimmer, new direction emerges", "skill1-1": "title: HCI Basic Knowledge | Learn HCI-related basic knowledge, such as human-centered design process, usability principles, and evaluation methods. This lays the foundation for your subsequent development in this direction.", "skill1-1-hint": "foundation is built", "skill1-2": "title: User Research | Acquire user research skills such as interviews, field studies, and usability testing. These skills enable you to better understand user needs and pain points, helping you design better interactive experiences for users.", "skill1-2-hint": "see through others' eyes", "skill1-3": "title: UI Prototyping | Learn how to use Figma, Sketch, ProtoPie, and other prototyping tools to create basic interactive prototypes. Mastering these tools allows you to quickly iterate on your design ideas in subsequent team projects.", "skill1-3-hint": "present ideas in reality" }"\\
\}
\end{flushleft}
\end{tcolorbox}

\begin{tcolorbox}[breakable, colback=gray!5, colframe=gray!75!black, title=Events Generation Prompt, fonttitle=\bfseries]
\ttfamily\footnotesize
\begin{flushleft}
\{ \\
\ \ \ {"instruction"}: "As a career planner, you can generate a major career development event that the user will encounter in the next four months, along with three random events, three skills they may learn, and corresponding information hints, based on the user's self-introduction, past experiences, and current situation. Output the results in JSON format.",\\

\ \ \ {"input"}: "\\
You are a career planner generating milestone<currentMilestoneNum>, three new random events, and three new skills based on:\\
User's profile All previously pocketed events Your output must be a strictly JSON formatted object containing: { "bigEvent<currentMilestoneNum>":"title: [Event Title] | [detailed description of the major career milestone that LOGICALLY FOLLOWS from all previous events]", "randomEvent<currentMilestoneNum>-1":"title: [Event Title] | [detailed event description that is DIRECTLY RELATED to bigEvent<currentMilestoneNum>, explaining what happens and its impact] [Positive/Neutral/Negative/Change: direction label]", "randomEvent<currentMilestoneNum>-1-hint":"[2-6 word extremely vague and ambiguous hint that does NOT reveal whether the event is positive, negative, or neutral]", "randomEvent<currentMilestoneNum>-2":"title: [Event Title] | [detailed event description that is DIRECTLY RELATED to bigEvent<currentMilestoneNum>, explaining what happens and its impact] [Positive/Neutral/Negative/Change: direction label]", "randomEvent<currentMilestoneNum>-2-hint":"[2-6 word extremely vague and ambiguous hint that does NOT reveal whether the event is positive, negative, or neutral]", "randomEvent<currentMilestoneNum>-3":"title: [Event Title] | [detailed event description that is DIRECTLY RELATED to bigEvent<currentMilestoneNum>, explaining what happens and its impact] [Positive/Neutral/Negative/Change: direction label]", "randomEvent<currentMilestoneNum>-3-hint":"[2-6 word extremely vague and ambiguous hint that does NOT reveal whether the event is positive, negative, or neutral]", "skill<currentMilestoneNum>-1":"title: [Skill Name] | [detailed description of this skill relevant to bigEvent<currentMilestoneNum>, what it involves, and how it helps current career stage]", "skill<currentMilestoneNum>-1-hint":"[2-6 word mysterious hint that doesn't clearly indicate what skill this is]", "skill<currentMilestoneNum>-2":"title: [Skill Name] | [detailed description of this skill relevant to bigEvent<currentMilestoneNum>, what it involves, and how it helps career progression]", "skill<currentMilestoneNum>-2-hint":"[2-6 word mysterious hint that doesn't clearly indicate what skill this is]", "skill<currentMilestoneNum>-3":"title: [Skill Name] | [detailed description of this skill relevant to bigEvent<currentMilestoneNum>, what it involves, and how it helps future development]", "skill<currentMilestoneNum>-3-hint":"[2-6 word mysterious hint that doesn't clearly indicate what skill this is]" }\\

Output requirements:\\
Format MUST STRICTLY be: "title: [title] | [detailed content description] [information label]" The format has THREE parts with specific structure: Title section: "title: [specific title]" Content section: " | [detailed content description]" Label section (for random events only): " [sentiment label]" Content descriptions MUST be detailed (at least 2-3 sentences) and cannot be omitted "bigEvent<currentMilestoneNum>" should be generated based on ALL pocketed events up to this point ALL random events and skills MUST be DIRECTLY RELATED to bigEvent<currentMilestoneNum> Skills should be specific abilities relevant to the current career stage with detailed descriptions Each randomEvent MUST end with label in square brackets: [Positive], [Neutral], [Negative], or [Change: direction] CRITICAL HINT REQUIREMENTS: -- ALL hints MUST be extremely vague, ambiguous, and mysterious -- NEVER use words that reveal the nature of the event or skill -- Use cryptic phrases like: "whispers emerge", "fog lifts slowly", "unknown beckons", "silence speaks", "doors appear", "winds shift" -- Hints should create curiosity without giving away whether something is beneficial or detrimental ALL bigEvents MUST occur within a TWO-YEAR timeframe from the start <acceptedChangesStr> NEVER omit any part of the required format Output only valid JSON User's profile: <userIntro> All pocketed events so far: <pocketedEvents> Past experiences: <pastExperiencesStr or 'None yet'> Accepted direction changes: <acceptedDirectionChanges> Current time: <currentTime>",\\

\}
\end{flushleft}
\end{tcolorbox}

\section{Appendix B: Prompts for Image and Career Report Generation}
\label{appendix:b}

\begin{tcolorbox}[breakable, colback=gray!5, colframe=gray!75!black, title=Milestone Image Generation Prompt, fonttitle=\bfseries]
\ttfamily\footnotesize
\begin{flushleft}
\{ \\

\ \ \ {"input"}: ""\$\{bigEventContent\}" Please identify the possible scenes in the above text and then generate a Ghibli-style image of that scene.Only the scene is needed, no text required."\\

\}
\end{flushleft}
\end{tcolorbox}

\begin{tcolorbox}[breakable, colback=gray!5, colframe=gray!75!black, title=Career Analysis Prompt, fonttitle=\bfseries]
\ttfamily\footnotesize
\begin{flushleft}
\{ \\
\ \ \ {"input"}: "Based on the user's 2-year career journey with collected milestones, events, and skills, generate a comprehensive career analysis and future suggestions.\\

Your output must be a strictly JSON formatted object containing: \{ "careerAnalysis":"(comprehensive analysis of the user's career journey, highlighting patterns, growth areas, skills acquired, and achievements. Include specific references to their journey events and decisions)", "futureSuggestions":"(specific, actionable suggestions for the next phase of their career based on their journey. Include 3-5 concrete next steps they should consider)" \}\\

User's profile: \$\{userIntro\}\\
Complete journey (all pocketed events): \$\{allEvents\}\\

Provide insightful analysis that:\\
- Identifies career trajectory patterns from their specific journey\\
- Recognizes strengths demonstrated through the events they experienced\\
- Highlights the skills they acquired and how to leverage them\\
- Acknowledges challenges they overcame\\
- Suggests concrete next steps based on their trajectory\\
- Offers encouragement for future growth\\
- References specific events from their journey\\

Output only valid JSON format."\\
\}
\end{flushleft}
\end{tcolorbox}

\section{Appendix C: Prompts for ChatGPT}
\label{appendix:c}

\begin{tcolorbox}[breakable, colback=gray!5, colframe=gray!75!black, title=ChatGPT Prompt, fonttitle=\bfseries]
\ttfamily\footnotesize
\begin{flushleft}
{ \
\ \ \ {"input"}: "You are a professional career development advisor. Based on my current background, please help me create a detailed two-year growth roadmap, including key milestones and actionable suggestions.\
\
My background is as follows:\
Current status (academic level or job position): (User Input)
My goal (where I want to be in two years—role, project, or achievement): (User Input))
\
Please provide systematic guidance from the following aspects:\
\

Stage-wise Goal Breakdown: How can I break down my two-year goal into phases (e.g., every 6 months or by semester)?\
\
Skills Development Checklist: What core skills should I focus on, and how should I develop each of them?\
\
Timeline of Key Milestones: Are there any critical deadlines (e.g., applications, competitions, hiring seasons, academic submissions) I should prepare for in advance?\
\
Common Pitfalls to Avoid: What are typical mistakes or traps in this path, and how can I proactively avoid them?\\}
\end{flushleft}
\end{tcolorbox}